\documentclass[%
 reprint,
showpacs,preprintnumbers,
 amsmath,amssymb,
 aps,
pra,
]{revtex4-1}

\usepackage{graphicx}
\usepackage{dcolumn}
\usepackage{bm}
\usepackage{epstopdf}
\usepackage{xcolor}

\begin{document}


\title{Structural rearrangements and fragmentation pathways induced by a low-energy electron attachment to ethyl acetate}
\author{Anirban Paul$^1$}
\author{Ian Carmichael$^2$}%
\author{Dhananjay Nandi$^{1,3}$}%
\author{Sylwia Ptasinska$^{2,4,\dagger}$}%
\author{Dipayan Chakraborty\textit{$^{2,\ast}$}}%
\affiliation{$^{1}$Indian Institute of Science Education and Research Kolkata, Mohanpur 741246, India}
\affiliation{$^{2}$Radiation Laboratory, University of Notre Dame, Notre Dame, Indiana 46556, United States.}
\affiliation{$^{3}$~Center for Atomic, Molecular and Optical Sciences $\&$ Technologies, Joint initiative of IIT Tirupati $\&$ IISER Tirupati, Yerpedu, 517619, Andhra Pradesh, India.}
\affiliation{$^{4}$~Department of Physics $\&$ Astronomy, University of Notre Dame, Notre Dame, IN 46556, USA}
\email{$^\dagger$sptasins@nd.edu}\email{$^\ast$dchakra2@nd.edu}



\date{\today}
\begin{abstract}
Exploring the molecular fragmentation dynamics induced by low-energy electrons offers compelling insights into the complex interplay between the projectile and target. In this study, we investigate the phenomenon of dissociative electron attachment to ethyl acetate. The recorded yields of various fragment anions within an incident electron energy range of 1 to 13 eV reveal a diverse array of products with six different mass numbers. Examples include  (M$-$H)$^-$, CH$_3^-$, C$_2$H$_5$O$^-$, CH$_3$CO$^-$, CH$_2$CHO$^-$, and CH$_3$COO$^-$, formed through the fracture of single bonds. Interestingly, the generation of other fragments, such as HCCO$^-$, suggests a more intricate structural rearrangement of the nuclei, adding a layer of complexity to the observed dissociation dynamics.
\end{abstract}

\maketitle

\section{Introduction}
Low-energy electron collisions with molecules are crucial processes in various scientific fields. They play a vital role in dissociation processes in the upper atmosphere and are also responsible for ozone layer depletion \cite{Lu2001}. Additionally, damage to living cells by nuclear radiation is primarily caused by low-energy secondary electrons, which are produced by high-energy primary beams through different processes such as ionization and multiple collisions. The single- and double-strand breaks of DNA are primarily caused by the impact of these low-energy secondary electrons through dissociative electron attachment (DEA) \cite{Boudaiffa}. DEA is the dominant process leading to the dissociation of molecules in low-energy electron collisions that have been experimentally observed. It is a two-step resonant process; in the first step, the incoming electron resonantly attaches to the molecule, producing a temporary negative ion (TNI). In the subsequent step, the TNI dissociates, producing an anionic fragment along with one or more neutral fragments \cite{illenberger2014gaseous,ptasinska_epjd}.\\
Ethyl acetate is among the simplest esters, a class of compounds derived from acids where the hydrogen atom of at least one acidic hydroxyl group ($-$OH) of that acid is replaced by an organyl group ($-$R). Esters are ubiquitous in nature and find extensive use in various industries. The pleasant aroma of many fruits, for instance, can be attributed to the presence of esters. Fats, on the other hand, are primarily triesters derived from glycerol and fatty acids, with glycerides representing fatty acid esters of glycerol and lactones being cyclic carboxylic esters. Ethyl acetate itself is a colorless liquid with a sweet, fruity odor that is generally well-received. It serves as a widely used solvent, particularly in the production of paints, varnishes, lacquers, cleaning mixtures, and perfumes \cite{Riemenschneider2005,buxton2001rate,Schneider2001}. Additionally, it is employed as a solvent in processes such as decaffeinating coffee beans and in column and thin-layer chromatography in laboratory settings \cite{Ramalakshmi1999,tan2003copper}. Furthermore, it serves as an asphyxiant for insect collection for study purposes. Industrially, ethyl acetate is primarily synthesized through the classic Fischer esterification reaction involving ethanol and acetic acid.\\
Due to its widespread usage in industry, ethyl acetate has been detected in the atmosphere, making it a potential source of CO$_2$ \cite{helmig1989volatile}. Therefore, it is important to study its decomposition in the atmosphere through the interaction with low-energy electrons. Despite its extensive use in research and industry, there has been no study on DEA to ethyl acetate to date \cite{Riemenschneider2005,buxton2001rate,Schneider2001,Ramalakshmi1999,tan2003copper}. Additionally, only a few reports exist on DEA to esters \cite{PARIAT1991181,Feketova2018,Illenberger2006}. Feketeov\'a \textit{et al.} \cite{Feketova2018} investigated the DEA dynamics of methyl acetate by measuring the yields of various fragment anions as a function of incident electron energies. They observed that these fragment anions were produced from one \emph{shape resonance} at approximately 2.5 eV, along with several \emph{Feshbach resonances} in the range of 6 - 12 eV electron energy. Pariat and Allan explored the DEA dynamics of methyl acetate by measuring the yields of different fragment anions \cite{PARIAT1991181}. They identified one \emph{shape resonance} near 3 eV and three higher energy \emph{Feshbach resonances} in the 6-12 eV range. Several anionic fragments were observed, including CH$_3^-$, CH$_3$CO$^-$, CH$_3$O$^-$, HCCO$^-$, CHO$^-$, CH$_2$COO$^-$ and CH$_2$COOCH$_3^-$ ions. They proposed a mechanism for the production of HCCO$^-$, involving the rapid stabilization of the primary resonance by the loss of an H-atom, followed by a slower process of proton transfer. The HCCO$^-$ anions produced due to low-energy electron collisions can act as a source of CO$_2$. These anions react with molecular oxygen, producing CO$_2$ \cite{PARIAT1991181}.
\section{Experimental and computational methods}
The experimental setup used to obtain ion-yield curves for negative ions has been previously detailed \cite{Chakraborty_NMF}. In the context of this present study, it will be briefly discussed. The experiment was conducted within an ultrahigh vacuum chamber with a base pressure of $1 \times 10^{-10}$ mbar, using a QMS by Hiden Analytical, specifically from the 3F series, commercially known as an ion desorption probe (IDP). In the present context, the system was employed for gas-phase detection of anions. An effusive beam of gas-phase molecules interacted with low-energy electrons generated by the oxide-coated iridium filament within the QMS. The filament emitted electrons through thermionic emission with controlled energy, typically at an intensity of 2.5 $\mu$A. When the low-energy electrons interacted with the molecules, negative fragment ions were produced and collected by a secondary electron multiplier detector with a selected mass-to-charge ratio (m/z). Energy scans were conducted from 1 to 13 eV, with 0.1 eV increments for each negative fragment generated through the DEA process. These energy scans were executed using MASsoft version 7 professional software provided by Hiden. The energy resolution of the electron beam used in the experiment was 0.5 eV. The chamber was baked at 363 K for several days before introducing the sample to reduce signals originating from contamination and background gases already present in the chamber. The electron energy scale was calibrated using the resonant peak of O$^-$ from DEA to O$_2$ \cite{rapp}. Energy calibration was performed both before and after conducting the experiment.
The experiments were carried out using 99.5$\%$ pure ethyl acetate samples from Sigma Aldrich, USA, which is liquid at room temperature. The sample was introduced from a glass container located outside the chamber and was transported to the interaction region through a pipeline connection. The pipeline was linked to an internal capillary within the vacuum chamber that guided the gas-phase sample to the QMS aperture. The sample underwent freeze-pump-thaw cycles several times to eliminate contaminants before being introduced into the vacuum chamber. The chamber pressure during the experiment was maintained at approximately 1.5$\times 10^{-6}$ mbar. The experiment was repeated multiple times for each fragment, and the results were consistent in each case.\\
Quantum chemical calculations were conducted using the GAUSSIAN 16 software \cite{g16}. Thermodynamic threshold energies for each dissociation channel were determined using the composite W1 method, \cite{W1_method} employing the B3LYP functional \cite{B3LYP} and the flexible cc-pVTZ basis set \cite{Dunning1989} for geometry optimizations and basic thermochemistry. These calculations were complemented by a series of correlation corrections\cite{Barnes_2009} to ensure high-level accuracy. The threshold energy for a specific dissociation channel of ethyl acetate was determined by calculating the bond dissociation energies involved and the electron affinity of the fragment.
\section{Results and discussions}
A total of six different mass fragments at 15, 41, 43, 45, 59, and 87 amu have been found due to the low-energy electron collision with ethyl acetate. It is important to note that six different mass numbers do not necessarily indicate the presence of six distinct fragment anions. Due to the complexity of the fragmentation process, anions with different molecular structures but with the same mass number are possible. The assignment of these masses to various fragment anions is provided in Table \ref{tbl:ResonanceAssignment}.\\
It is important to note that the production of the fragments CH$_3^-$, CH$_3$CO$^-$, C$_2$H$_5$O$^-$, CH$_3$COO$^-$, and the H-loss anions [(M$-$H)$^-$] can be attributed to single bond breaking. In contrast, the generation of HCCO$^-$ and CH$_2$CHO$^-$ involves multiple bond breaking and/or complex structural rearrangement. The absence of C$_2$H$_5^-$ ions can be explained by the negative electron affinity of C$_2$H$_5$. For reference, the electron affinities of the neutral counterparts of several fragment anions are provided in Table \ref{tbl:EAs}. \\
In the following sections, we will first examine the ions generated through single bond breaking or direct dissociation. Subsequently, we will delve into the production of ions that necessitate complex structural rearrangements. We obtained the negative-ion yields of different fragment anions produced from DEA to ethyl acetate as a function of incident electron energy ranging from 1 to 13 eV, and the resulting fragment anions, along with their respective assignments and resonance positions, are provided in Table \ref{tbl:ResonanceAssignment}.
\subsection{Production of H-loss anions [(M$-$H)$^-$] (M = 87)} \label{M-H}
\begin{figure}[h]
\centering
  \includegraphics[scale=0.56]{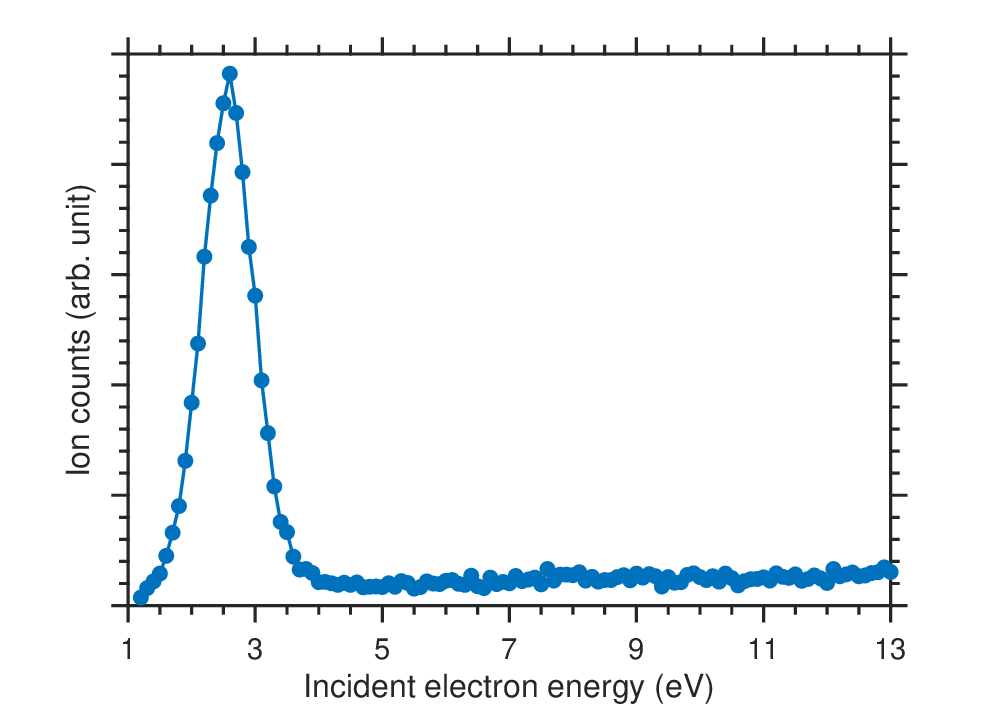}
  \caption{Ion yield curve of (M$-$H)$^-$ ions produced due to DEA to ethyl acetate in the energy range of 1-13 eV. A single resonant peak at 2.7 eV is observed.}
  \label{fig:ion_yield_M_H}
\end{figure}

The yield of (M$-$H)$^-$ ions, as a function of electron energy during DEA to ethyl acetate, is depicted in Fig. \ref{fig:ion_yield_M_H}. This ion yield exhibits a prominent peak centered around 2.7 eV that can be attributed to resonance with a possible $\pi^*$ character. The previous photoabsorption study on ethyl acetate indicates that the lowest unoccupied molecular orbital (LUMO) predominantly exhibits $\pi^*$ antibonding character and is localized on the C$=$O bond \cite{Paulo2016}. Consequently, during the attachment, the incoming electron can be captured by the unoccupied $\pi^*$ (C$=$O) orbital, leading to the formation of the TNI state, which subsequently dissociates, producing various fragment anions. In the current investigation, the observed resonance at 2.7 eV is attributed to this $\pi^*$ \emph{shape resonance}. In a previous study of DEA to methyl acetate, a similar resonance in this energy range was also identified and attributed to a $\pi^*$ resonance \cite{PARIAT1991181}. In the context of the TNI state, three distinct dissociation channels leading to fragment anions with the same mass number but different molecular structures are possible, depending on the site of H-loss. These three H-loss anions are CH$_2$COOCH$_2$CH$_3^-$, CH$_3$COOCH$_2$CH$_2^-$, and CH$_3$COOCHCH$_3^-$. The first one is formed by breaking a C$-$H bond from the methyl group. In contrast, the second and third ones are generated by breaking a C$-$H bond within the CH$_3$ and CH$_2$ groups, respectively, of the ethyl moiety. These ions are exclusively produced through the cleavage of a single C$-$H bond. The possible dissociation channels responsible for producing these H-loss anions can be represented as follows:
\begin{table}
  \centering
  \caption{Resonant positions of the fragment anions obtained in DEA to ethyl acetate.}
    \begin{tabular}{clcc}
    \hline
    \multicolumn{1}{c}{Mass} & \multicolumn{1}{c}{Assignment} & \multicolumn{2}{c}{Peak position (eV)} \\
    (Da) & \multicolumn{1}{c}{of ions} & Shape & Feshbach \\
         &  & Resonance & Resonance \\
    \hline
    15     & CH$_3^-$ & - & 5.7, 7.3, 8.5, 10.1 \\
    \hline
    41     & HCCO$^-$ & 3.3 & 7.0, 9.7 \\
    \hline
    43     & CH$_3$CO$^-$ $\&$ CH$_2$CHO$^-$ & - & 7.3, 8.2, 10.1 \\
    \hline
    45     & CH$_3$CH$_2$O$^-$ & 3.0 & 7.3, 9.9 \\
    \hline
    59     & CH$_3$COO$^-$ & 3.0 & 8.9, 10.2 \\
    \hline
        & CH$_2$COOCH$_2$CH$_3^-$ &  &  \\
    87    &  CH$_3$COOCH$_2$CH$_2^-$ & 2.7 & - \\
        &  CH$_3$COOCHCH$_3^-$ &  &  \\
    \hline
    \end{tabular}
  \label{tbl:ResonanceAssignment}
\end{table}
\begin{table}
\centering
\caption{Electron affinities (EA) of selected neutrals.}
\begin{tabular} {lc}
\hline
Species & Electron affinities (EA)\\
\hline
CH$_3$  & 0.029 eV \\

C$_2$H$_5$  & -0.236 eV \\

CH$_3$CO & 0.403 eV \\

CH$_2$CHO & 1.837 eV \\

HCCO  & 2.353 eV \\

C$_2$H$_5$O  & 1.678 eV \\

CH$_3$COO & 3.328 eV \\

CH$_3$COOCH$_2$ & 0.478 eV \\

CH$_2$COOC$_2$H$_5$  & 1.692 eV \\

CH$_3$COOCHCH$_3$  & 0.355 eV \\
\hline
\end{tabular}
\label{tbl:EAs}
\end{table}

\begin{align*}
(\mbox{C}\mbox{H}_{3} \mbox{COO}\mbox{C}_{2}\mbox{H}_{5})^{*-} & \rightarrow \mbox{C}\mbox{H}_{2} \mbox{COO}\mbox{C}\mbox{H}_{2}\mbox{C}\mbox{H}_{3}^{-} + \mbox{H} \tag{1a} \\
& \rightarrow \mbox{C}\mbox{H}_{3}\mbox{COO}\mbox{C}\mbox{H}_{2}\mbox{C}\mbox{H}_{2}^{-} + \mbox{H} \tag{1b} \\
& \rightarrow \mbox{C}\mbox{H}_{3}\mbox{COO}\mbox{C}\mbox{H}\mbox{C}\mbox{H}_{3}^{-} + \mbox{H} \tag{1c} 
\end{align*}

Our theoretical investigation indicates that CH$_3$COOCH$_2$CH$_2^-$ ions in channel (1b) are inherently unstable and undergo spontaneous dissociation, yielding a CH$_3$COO$^-$ ion along with a neutral conjugate of CH$_2$CH$_2$ (Channel 5b). Consequently, in the current context, channel (1b) can be disregarded for the generation of (M$-$H)$^-$ ions. The calculated threshold energies for channels (1a) and (1c) are 2.55 and 3.83 eV, respectively.\\

As we investigated the H-loss channels, the anticipation was to observe conjugate dissociation channels involving the formation of H$^-$ ions in this energy range. However, our current experimental limitations prevent us from verifying this expectation. The QMS spectrometer used in this experiment lacks the capability to detect H$^-$ ions.
\begin{figure}[h]
\centering
  \includegraphics[scale=0.52]{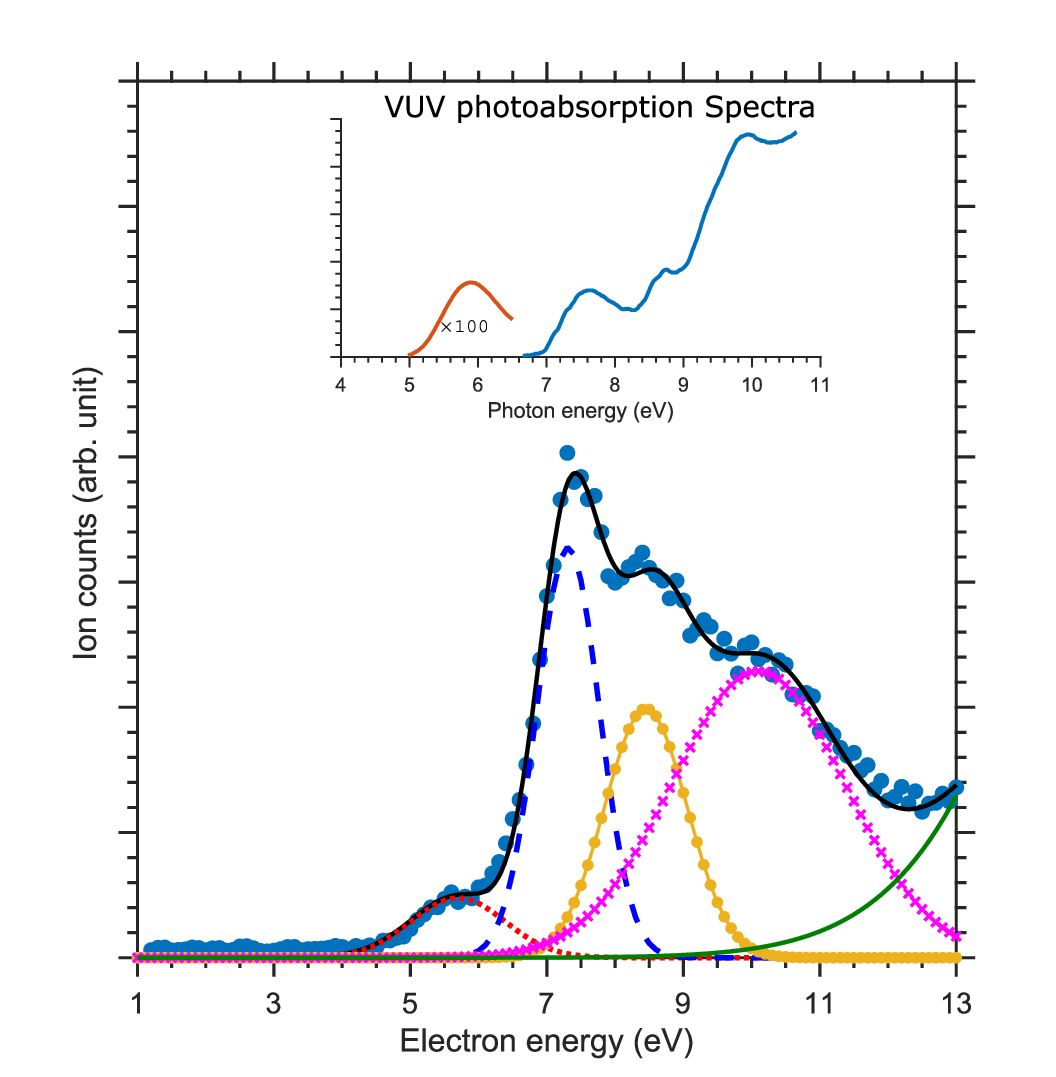}
  \caption{The VUV photoabsorption spectrum of ethyl acetate was extracted from the previous report (CC BY) \cite{Paulo2016}. This spectrum reveals the presence of four absorption bands within the 5-10 eV energy region. The ion yield of CH$_3^-$ ions produced due to DEA to ethyl acetate in the energy range of 1-13 eV is also presented at the bottom. The blue dots represent the experimental data points, and the black solid curve indicates the cumulative fitted curve with four Gaussian functions attributed to four resonant bands. One exponential function is also shown to incorporate the contribution of the IP dissociation process. The fitting suggests that the structure present in the 5-12 eV range is a combination of four resonant states with energies at 5.7, 7.3, 8.5, and 10.1 eV.}
  \label{fig:ion_yield_CH3}
\end{figure}
\subsection{Production of $\mbox{CH}_{3}^{-}$ ions (M = 15)} \label{CH3}
Fig. \ref{fig:ion_yield_CH3} illustrates the ion yield curve of CH$_3^-$ ions resulting from DEA to ethyl acetate. A broad range of overlapping resonances, spanning from 6 eV to 12 eV, is observed, accompanied by a hump near 5.7 eV. To locate the correct positions of these overlapping resonances, we opted to fit the CH$_3^-$ ion yield using multiple Gaussian functions, as depicted in Fig. \ref{fig:ion_yield_CH3}. The fitted ion yield curve indicates the likely presence of three closely situated resonant states in the 6 to 12 eV region, peaking at 7.3, 8.5, and 10.1 eV, respectively. However, due to the limited energy resolution of the electron beam and the finite width of the resonances, it is not feasible to distinctly separate them in the present study.\\
In order to have an insight into these different resonances, we compare them to the vacuum ultraviolet (VUV) photoabsorption study \cite{Paulo2016}. The study suggests the existence of an absorption band in the vicinity of 5.9 eV, attributed to the HOMO $\rightarrow \pi^*$(C$=$O) transition. Additionally, the authors anticipate the occurrence of multiple Rydberg transitions spanning the 6.5-10 eV range, as shown in Fig. \ref{fig:ion_yield_CH3}. These Rydberg states are proposed to serve as parent states for the DEA resonances observed in the present investigation. Therefore, through comparison with the prior report, the four DEA resonances identified within the 5 to 12 eV energy range are likely to be \emph{core-excited Feshbach} resonances. The first band in the photoabsorption spectra, which peaks at approximately 5.9 eV, may serve as the parent state for the weak resonance detected at 5.7 eV. Likewise, the second band, reaching its peak at about 7.7 eV, could be the parent state for the sharp resonance identified at 7.3 eV. Similarly, the remaining two bands, with peaks around 8.8 eV and 9.8 eV, respectively, may correspond to the parent states for resonances at 8.5 eV and 10.1 eV. A comparison between the VUV photoabsorption spectrum \cite{Paulo2016} and the CH$_3^-$ ion yield is depicted in Fig. \ref{fig:ion_yield_CH3}.\\
The CH$_3^-$ formation from ethyl acetate could be through a simple two-body dissociation process either from the methyl site or from the ethyl moiety, as shown by the following reaction channels.
\begin{align*}
(\mbox{C}\mbox{H}_{3}\mbox{COO}\mbox{C}_{2}\mbox{H}_{5})^{*-} 
& \rightarrow \mbox{C}\mbox{H}_{3}^{-} + \mbox{COO}\mbox{C}_{2}\mbox{H}_{5} \tag{2a} \\
& \rightarrow \mbox{C}\mbox{H}_{3}^{-} + \mbox{C}\mbox{H}_{3}\mbox{COO}\mbox{C}\mbox{H}_{2} \tag{2b} 
\end{align*}
In channel (2a), CH$_3^-$ ions are produced from the $-$CH$_3$ site, while in channel (2b), CH$_3^-$ ions are produced from the $-$C$_2$H$_5$ moiety. The theoretically calculated threshold values for these channels are 4.01 and 3.81 eV, respectively, rendering both channels relevant to the DEA process. However, it is intriguing to explore whether three-body or even higher-order dissociation of the TNI is possible. To be transparent, the present measurements do not provide a conclusive answer to this question. Nevertheless, by examining the resonance positions, we can speculate that the 5.7 eV resonance primarily involves two-body dissociation, whereas the 7.3, 8.5, and 10.1 eV resonances may either undergo two-body dissociation with significantly high rovibrational excited fragments or involve three-body or higher-order dissociations. The higher values of excess available energy in the dissociation process contribute to this complexity. To offer a definitive answer, measuring the kinetic energy of the CH$_3^-$ fragments would be necessary, which goes beyond the scope of the present study.\\
\begin{table}[h!]
  \centering
    \caption{\label{tab:DEA-Channels}Dissociation channels producing different fragment anions and their computed threshold values.}
    
    \begin{tabular}{clc}
    \hline
    Ch. no. & Dissociative products & Threshold energy (eV) \\
    \hline
    (1a)     & CH$_3$COOCH$_2$CH$_2^-$ + H & 2.55 \\
    \hline
    (1c)     & CH$_3$COOCHCH$_3^-$ + H & 3.83 \\
    \hline
    & & \\
    \hline
    (2a)     & CH$_3^-$ + COOC$_2$H$_5$ & 4.01 \\
    \hline
    (2b)     & CH$_3^-$ + CH$_2$COOCH$_2$ & 3.81 \\
    \hline
    (2c)     & CH$_3^-$ + COOC$_2$H$_5^+$ & 11.13 \\
    \hline
    (2d)     & CH$_3^-$ + CH$_2$COOCH$_2^+$ & 11.15 \\
    \hline
     & & \\
    \hline
    (3a)     & CH$_2$CH$_3$O$^-$ + CH$_3$CO & 2.62 \\
    \hline
    (3b)     & CH$_2$CH$_3$O$^-$ + CH$_3$ + CO & 3.04 \\
    \hline
     & & \\
    \hline
    (4a)     & CH$_3$CO$^-$ + C$_2$H$_5$O & 3.90 \\
    \hline
    (4b)     & CH$_3$CO + CH$_2$CHO$^-$ + H$_2$ & 2.74 \\
    \hline
     & & \\
    \hline
    (5a)     & CH$_3$COO$^-$ + C$_2$H$_5$ & 0.57 \\
    \hline
     (5b)   &  CH$_3$COO$^- +$ CH$_2$CH$_2 +$ H & 2.31 \\
    \hline
     & & \\
    \hline
    (6a)     & HCCO$^-$ + C$_2$H$_5$O  + H$_2$ & 3.85 \\
    \hline
    (6b)     & HCCO$^-$  + CH$_3$CO  + 2 H$_2$ & 3.85 \\
    \hline
    (6c)     & HCCO$^-$ + C$_2$H$_5$  + H$_2$O & 2.73 \\
    \hline
    \end{tabular}
\end{table}
In this context, it is worth mentioning that if the production of CH$_3^-$ occurs due to the breaking of the H$_3$C$-$COOC$_2$H$_5$ bond, there should also be a possibility of producing COOC$_2$H$_5^-$ ions. Conversely, if CH$_3^-$ ions are produced due to the breaking of the CH$_3$COOCH$_2-$CH$_3$ bond, there should be a chance of producing CH$_3$COOCH$_2^-$ ions. Both COOC$_2$H$_5^-$ and CH$_3$COOCH$_2^-$ have a mass of 73, which is absent from our mass spectra. Our theoretical calculations suggest that the former ion spontaneously dissociates to produce one C$_2$H$_5$O$^-$ ion and one CO molecule (Channel 3b). However, CH$_3$COOCH$_2^-$ is a stable species with a positive electron affinity. The computed threshold energy for the production of this anion is 3.36 eV. The absence of CH$_3$COOCH$_2^-$ in our results may be due to either the CH$_3$COOCH$_2-$CH$_3$ bond not breaking to produce CH$_3^-$ ions or if the CH$_3$COOCH$_2-$CH$_3$ bond is breaking, CH$_3$COOCH$_2^-$ is not being produced. However, electron affinity calculations suggest that the electron affinity of CH$_3$ (0.029 eV) is lower than that of CH$_3$COOCH$_2$ (0.478 eV). This implies that the probability of producing CH$_3$COOCH$_2^-$ should be higher than that of CH$_3^-$. This observation likely indicates the non-dissociative nature of the CH$_3$COOCH$_2-$CH$_3$ bond upon electron attachment to ethyl acetate. Hence, the observed CH$_3^-$ ions are coming from the H$_3$C$-$COOC$_2$H$_5$ dissociation (Channel 2a). A similar observation was reported in a previous study on DEA of methyl acetate molecules \cite{PARIAT1991181}. By comparing the ion yields with deuterated methyl acetate, the authors experimentally demonstrated that the formation of CH$_3^-$ is favored through dissociation from the H$_3$C$-$COOCH$_3$ site.\\

A continuous increase in ion counts with increasing incident electron energy is observed above 11 eV. This behavior is due to the involvement of the CH$_3^-$ ions formed due to the ion pair (IP) dissociation \cite{ChakrabortyIPD,Chakraborty_2023}. For other fragment anions, the threshold of the IP states lies above 14 eV, hence absent in the measured ion yield. However, that is not the case for the CH$_3^-$ ions, where the calculated IP threshold lies around 11.1 eV. The CH$_3^-$ ion yield feature at higher energies agrees with our theoretically calculated IP threshold. It is worth investigating why the IP states present near 11.1 eV dissociate through producing the CH$_3^-$ ions only. At the present moment, we do not have any explanation behind this. Another striking observation is the presence of IP states at such low energy, which is not so common and observed only for a few investigations until now \cite{Chakraborty_CH2F2}. Two possible channels producing CH$_3^-$ ions through the IP dissociation are listed below:
\begin{align*}
e^- + \mbox{C}\mbox{H}_{3}\mbox{COO}\mbox{C}_{2}\mbox{H}_{5} & \rightarrow \mbox{C}\mbox{H}_{3}^{-} + \mbox{COO}\mbox{C}_{2}\mbox{H}_{5}^{+} + e^- \tag{2c} \\
& \rightarrow \mbox{C}\mbox{H}_{3}^{-} + \mbox{C}\mbox{H}_{3}\mbox{COO}\mbox{C}\mbox{H}_{2}^{+} + e^- \tag{2d} 
\end{align*}
The theoretically computed threshold values of these two channels are 11.13 and 11.15 eV, respectively. The overlap between the DEA and the IP dissociation is observed from the CH$_3^-$ ion yield. In order to include the contribution from the IP dissociation into the ion yield, we fitted it with an exponential function (along with the Gaussian functions) as shown in Fig. \ref{fig:ion_yield_CH3}.
\subsection{Production of C$_2$H$_5$O$^-$ ions (M = 45)}
\begin{figure}
\centering
  \includegraphics[scale=0.56]{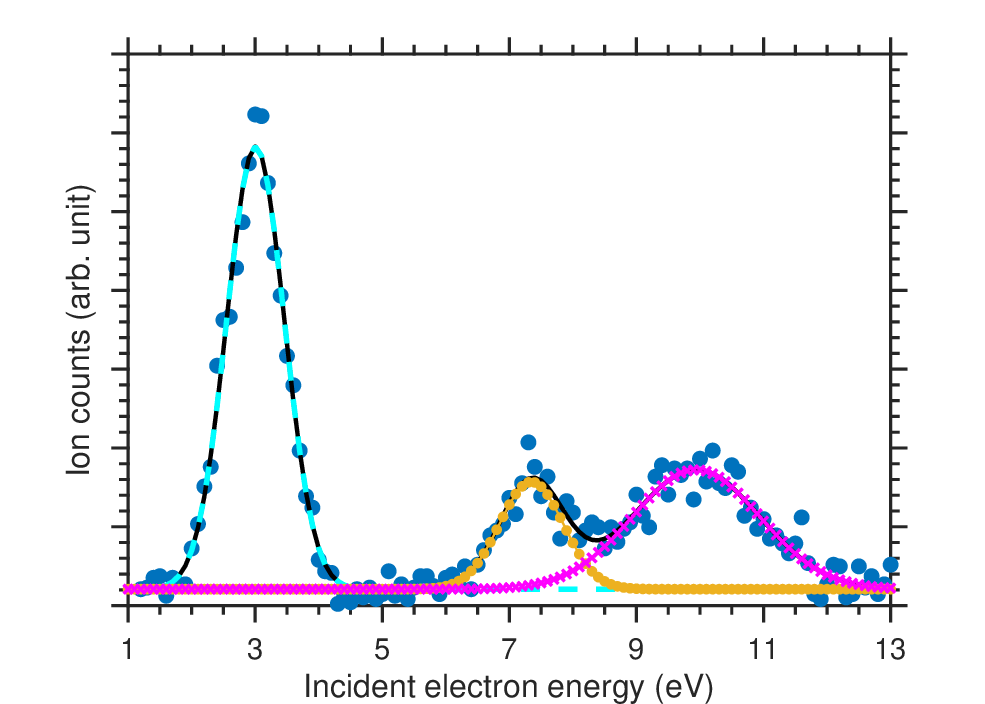}
  \caption{The ion yield of C$_2$H$_5$O$^-$ ions produced due to DEA to ethyl acetate in the energy range of 1 to 13 eV. The blue dots represent the experimental data points, and the black solid line represents the cumulative fit of the ion yield with four Gaussian functions. The fitting suggests that the structure present in the 3.0 eV region is due to the contribution of a \emph{shape resonance}, whereas the 6-12 eV region is a combination of two \emph{core-excited Feshbach} resonances peaking at 7.3, and 9.9 eV.}
  \label{fig:ion_yield_C2H5O}
\end{figure}

Fig. \ref{fig:ion_yield_C2H5O} illustrates the ion yield of C$_2$H$_5$O$^-$ resulting from DEA to ethyl acetate. A distinct and intense peak is observed at 3 eV, accompanied by two broad resonances peaking at 7.3 and 9.9 eV. Due to the broad nature of these resonant states, the contribution of other resonances with a lower cross-section cannot be ruled out in this energy region. The most straightforward mechanism for the formation of C$_2$H$_5$O$^-$ ions during the DEA process is through the dissociation of the CH$_3$OC$-$OC$_2$H$_5$ bond. This simple two-body dissociation can be represented as follows:
\begin{align*} 
& (\mbox{C}\mbox{H}_{3}\mbox{COO}\mbox{C}_{2}\mbox{H}_{5})^{*-} \rightarrow \mbox{C}\mbox{H}_{3}\mbox{CO} + \mbox{C}_{2}\mbox{H}_{5}\mbox{O}^{^-} \tag{3a} 
\end{align*}
The calculated threshold energy for this channel is 2.62 eV, making it the only possible candidate for the observed 2.9 eV \emph{shape resonance}. The experimentally obtained appearance energy aligns well with the computed threshold energy, considering the resolution of the electron beam used in the experiment. Regarding the \emph{Feshbach resonances} within the 6 to 12 eV region, three-body or even higher-order dissociation might be involved. One such possible sequential dissociation channel is:
\begin{align*}
(\mbox{C}\mbox{H}_{3}\mbox{COO}\mbox{C}_{2}\mbox{H}_{5})^{*-} & \rightarrow \mbox{C}\mbox{H}_{3}+ \mbox{COO} \mbox{C}_{2}\mbox{H}_{5}^{^-} \\
& \rightarrow \mbox{C}\mbox{H}_{3}+ \mbox{CO} + \mbox{C}_{2}\mbox{H}_{5}\mbox{O}^{^-}\tag{3b} 
\end{align*}
The calculated threshold energy for this channel is 3.04 eV, making it a potential candidate for the \emph{Feshbach resonances}. This proposed dissociation channel (3b) could either be a concerted three-body dissociation or a sequential dissociation. In order to make this dissociation channel more relevant, one can relate it to the CH$_3^-$ dissociation channel as discussed in section \ref{CH3}. The dissociation of the H$_3$C$-$COOC$_2$H$_5$ bond produces CH$_3^-$ ions with a neutral COOC$_2$H$_5$ conjugate. This introduces the possibility of observing COOC$_2$H$_5^-$ ions in the DEA process. However, we are unable to detect the presence of COOC$_2$H$_5^-$ in our measurements. One of the reasons for this could be its unstable nature or its short lifetime. Therefore, it is plausible that COOC$_2$H$_5^-$ ions are indeed formed in the DEA process, but due to their unstable nature, they spontaneously dissociate to produce CO and C$_2$H$_5$O$^-$ ions. The presence of \emph{Feshbach resonances} within the 6 to 12 eV region for the CH$_3^-$ ions reinforces this conclusion. This observation suggests that a sequential dissociation of the TNI may involved.
\subsection{Production of CH$_3$CO$^-$ and CH$_2$CHO$^-$ ions (M = 43)}
\begin{figure}
\centering
  \includegraphics[scale=0.56]{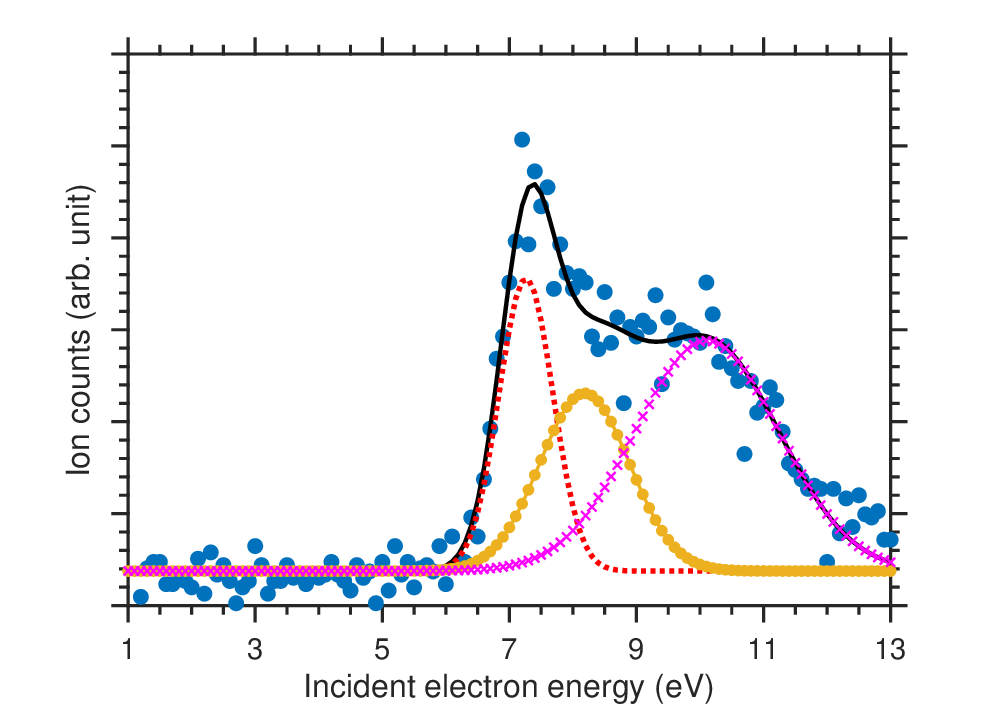}
  \caption{The ion yield of CH$_3$CO$^-$ ions produced due to DEA to ethyl acetate in the energy range of 1 to 13 eV. The blue dots represent the experimental data points, and the black solid line represents the cumulative fit of the ion yield with three Gaussian functions. The fitting suggests that the structure present in the 6-12 eV region is a combination of three \emph{core-excited Feshbach} resonances peaking at 7.3, 8.2, and 10.1 eV.}
  \label{fig:ion_yield_CH3CO}
\end{figure}

Fig. \ref{fig:ion_yield_CH3CO} depicts the ion yield of mass 43 produced due to low-energy electron attachment to ethyl acetate. The presence of \emph{Feshbach resonances} within the 6 to 12 eV region is detected. However, no contribution from the \emph{shape resonance} is observed for fragments with this mass. The easiest way to form mass 43 is through a two-body dissociation by breaking the CH$_3$OC$-$OC$_2$H$_5$ bond, resulting in CH$_3$CO$^-$ ions with C$_2$H$_5$O as the neutral. The dissociation pathway for this fragment is as follows:
\begin{align*}
& (\mbox{C}\mbox{H}_{3}\mbox{COO}\mbox{C}_{2}\mbox{H}_{5})^{*-} \rightarrow \mbox{C}\mbox{H}_{3}\mbox{CO}{^-} + \mbox{C}_{2}\mbox{H}_{5}\mbox{O} \tag{4a} 
\end{align*}
This is essentially the conjugate dissociation process of C$_2$H$_5$O$^-$ formation (channel 3a), where we observed a strong signal due to the contribution from the \emph{shape resonance} within the 1 to 4 eV energy region (Fig. \ref{fig:ion_yield_C2H5O}). However, no ion signal within the 1 to 4 eV region is observed in the CH$_3$CO$^-$ channel. The higher threshold energy of this dissociation channel is the reason behind the exclusion of the \emph{shape resonance}, as we found the computed threshold energy to be around 3.9 eV.\\
Another possible contribution for mass 43 could be due to CH$_2$CHO$^-$, which requires many-body dissociation along with structural rearrangements in the TNI. This dissociation initiates with the production of C$_2$H$_5$O$^-$, which simultaneously or sequentially dissociates, producing CH$_2$CHO$^-$ and H$_2$. The presence of C$_2$H$_5$O$^-$ ions within the 6 to 12 eV \emph{Feshbach resonance} region (Fig. \ref{fig:ion_yield_C2H5O}) makes this a candidate dissociation pathway. The only conceivable reason to exclude this channel could be the stability of C$_2$H$_5$O$^-$ ions. However, studies on DEA to ethanol indicate a finite possibility behind the dissociation of C$_2$H$_5$O$^-$ ions to produce a CH$_2$CHO$^-$ ion and molecular Hydrogen (H$_2$) \cite{Ibanescu2007,Anirban_Ethanol}.
This process can be represented as:
\begin{align*}
(\mbox{C}\mbox{H}_{3}\mbox{COO}\mbox{C}_{2}\mbox{H}_{5})^{*-} & \rightarrow \mbox{C}\mbox{H}_{3}\mbox{CO} + \mbox{C}_{2}\mbox{H}_{5}\mbox{O}^{^-} \\ 
& \rightarrow \mbox{C}\mbox{H}_{3} \mbox{CO} + \mbox{C}\mbox{H}_{2}\mbox{CHO}^{^-} + \mbox{H}_{2} \tag{4b} 
\end{align*}
The computed threshold energy of this channel is 2.74, which is in agreement with the present experimental results. The significant difference between the computed threshold energy and the experimentally observed appearance energy for this channel implies that the resulting fragments will either exhibit high kinetic energy or exist in significantly elevated rovibrational excited states. Additionally, the possibility of further sequential dissociation of the neutrals cannot be ruled out.
\subsection{Production of CH$_3$COO$^-$ ions (M = 59)}
\begin{figure}
\centering
  \includegraphics[scale=0.56]{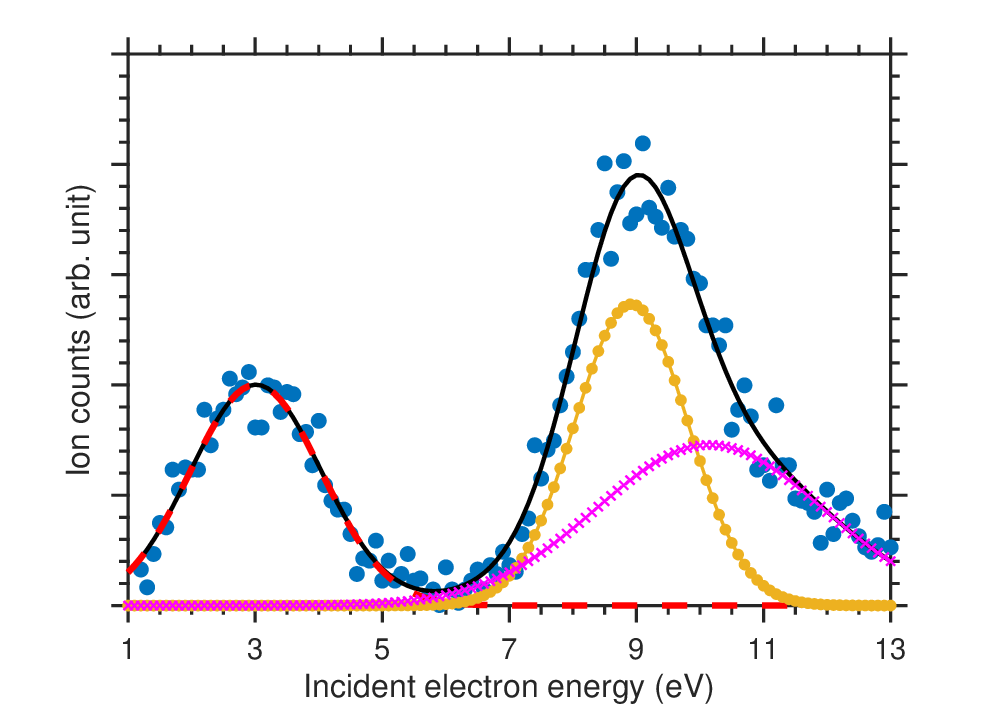}
  \caption{The ion yield of CH$_3$COO$^-$ ions produced due to DEA to ethyl acetate in the energy range of 1 to 13 eV. The blue dots represent the experimental data points, and the black solid line represents the cumulative fit of the ion yield with four Gaussian functions. The fitting suggests that the structure present in the 3 eV region is due to the contribution of a \emph{shape resonance}, whereas the 6-12 eV region is a combination of two \emph{core-excited Feshbach} resonances peaking at 8.9, and 10.2 eV.}
  \label{fig:ion_yield_CH3COO}
\end{figure}

The yield of CH$_3$COO$^-$ ions as a function of the electron energy from DEA to ethyl acetate is depicted in Fig. \ref{fig:ion_yield_CH3COO}. Two broad resonances peaking at 3 eV and 9 eV are observed. The higher energy resonant feature ranges from 6 to 12 eV, indicating the presence of similar \emph{Feshbach resonances} observed for other fragments. The simplest way to form the CH$_3$COO$^-$ ion is through a two-body dissociation channel by breaking the CH$_3$COO$-$C$_2$H$_5$ bond. The dissociation channel can be represented as follows:
\begin{align*}
(\mbox{C}\mbox{H}_{3}\mbox{COO}\mbox{C}_{2}\mbox{H}_{5})^{*-} 
& \rightarrow \mbox{C}\mbox{H}_{3}\mbox{COO}^{-} + \mbox{C}_{2}\mbox{H}_{5} \tag{5a}  
\end{align*}
The computed threshold energy for this dissociation channel is 0.57 eV, making this channel plausible for the 3 eV resonance. For the higher resonances, three-body or higher-order dissociation may be involved. 
In this context, it is worth mentioning that since the production of CH$_3$COO$^-$ occurs due to the breaking of the CH$_3$COO$-$C$_2$H$_5$ bond, there should also be a possibility of the production of C$_2$H$_5^-$ ions. However, in the present context, we could not detect any signal for mass 29. Our theoretical calculation suggests that C$_2$H$_5$ has negative electron affinity, and thus, the production of C$_2$H$_5^-$ is not possible in this context.\\
Another possible dissociation channel for the formation of the CH$_3$COO$^-$ ions is through a sequential dissociation of the TNI. As mentioned in section \ref{M-H}, CH$_3$COOCH$_2$CH$_2^-$ (channel 1b) ion is extremely unstable and simultaneously dissociates to produce a CH$_3$COO$^-$ ion with a CH$_2$CH$_2$ neutral conjugate. The dissociation mechanism can be represented in the following way:
\begin{align*}
(\mbox{C}\mbox{H}_{3}\mbox{COO}\mbox{C}_{2}\mbox{H}_{5})^{*-} 
 & \rightarrow \mbox{C}\mbox{H}_{3}\mbox{COO}\mbox{CH}_{2}\mbox{CH}_{2}^{-} + \mbox{H} \\
& \rightarrow \mbox{C}\mbox{H}_{3}\mbox{COO}^{-} + \mbox{CH}_{2}\mbox{CH}_{2} + \mbox{H}\tag{5b}
\end{align*}
The computed threshold energy for this dissociation channel is 2.31 eV, making it a candidate for both lower and higher energy resonances.
\begin{figure}
\centering
  \includegraphics[scale=0.56]{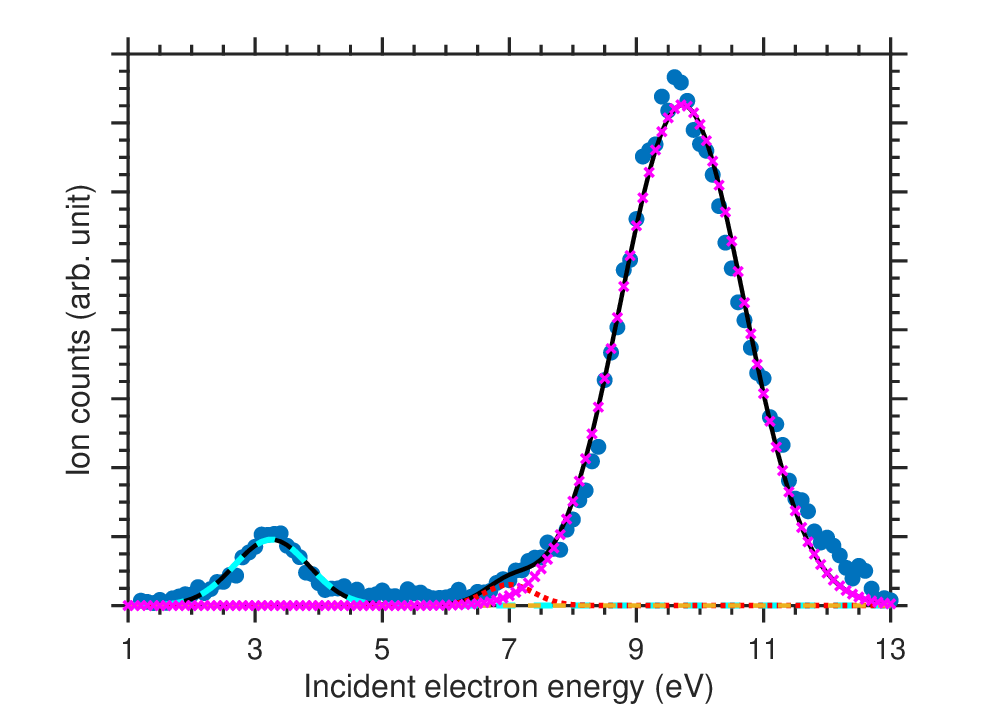}
  \caption{The ion yield of HCCO$^-$ ions produced due to DEA to ethyl acetate in the energy range of 1 to 13 eV. The blue dots represent the experimental data points, and the black solid line represents the cumulative fit of the ion yield with three Gaussian functions. The fitting suggests that the structure present in the 3.3 eV region is due to the contribution of a \emph{shape resonance}, whereas the 6-12 eV region is a combination of two \emph{core-excited Feshbach} resonances peaking at 7, 9.7 eV.}
  \label{fig:ion_yield_CHCO}
\end{figure}
\subsection{Production of HCCO$^-$ ions (M = 41)}
Fig. \ref{fig:ion_yield_CHCO} illustrates the ion yield curve of mass 41 resulting from low-energy electron attachment to ethyl acetate. One small peak near 3.3 eV and a broad peak near 9.7 eV are observed, along with a small hump near 7 eV. To clearly locate these resonances, we opted to fit the ion yield with three Gaussian functions, as shown in Fig. \ref{fig:ion_yield_CHCO}. Mass 41 can be attributed to the formation of HCCO$^-$ ions. This is likely the only fragment that cannot be produced through a single-bond dissociation process; instead, complex structural rearrangements in the TNI are essential. Various dissociation channels producing HCCO$^-$ ions are listed in Table \ref{tab:DEA-Channels}. In all these channels, complex structural changes and hydrogen migration are involved \cite{YU2020123}. To elucidate this process, we can refer to Fig. \ref{fig:CHCO_Production}. The formation of HCCO$^-$ fragments can proceed either through C$-$O dissociation near the keto site or near the ethyl site. We are discussing the processes as follows:
\begin{figure*}
\centering
  \includegraphics[scale=0.28]{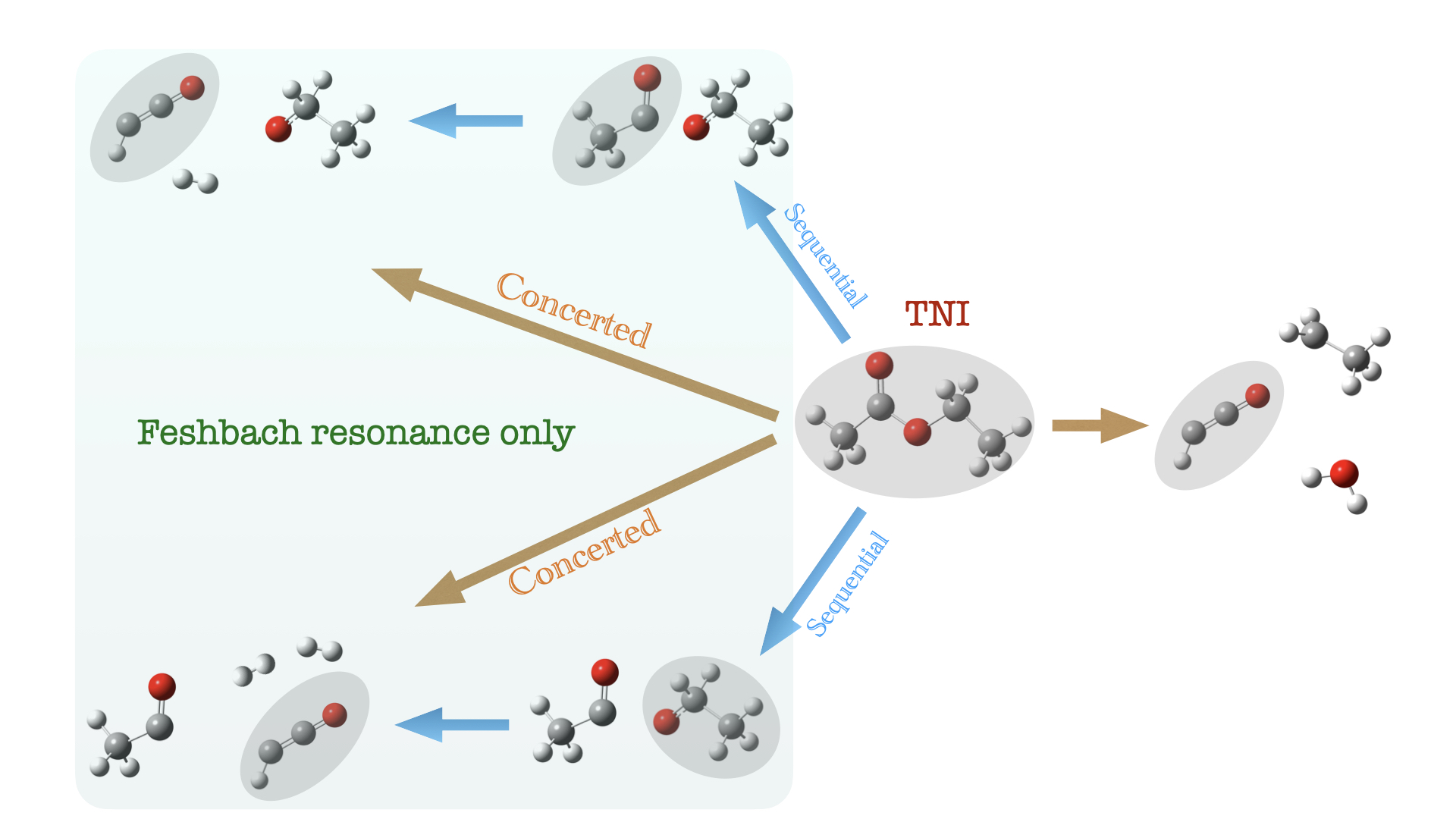}
  \caption{Suggested fragmentation pathways of ethyl acetate producing HCCO$^-$ during the DEA process. Structural rearrangements in the molecular anion must occur. The two fragmentation channels on the left show the pathways when the C$-$O dissociation occurred near the keto site of the molecule, whereas the right pathway represents when the C$-$O dissociation occurs near the ethyl group.}
  \label{fig:CHCO_Production}
\end{figure*}
\subsubsection{C$-$O dissociation near the keto site}
In this dissociation channel, molecular H$_2$ forms as a by-product. The dissociation can occur through a concerted three-body process involving structural rearrangements in the TNI. One C$-$O bond dissociates, two C$-$H bonds dissociate from the CH$_3$C$=$O site, and a H$-$H bond forms during this process. The final dissociated products of this channel are shown in channel (6a). Alternatively, HCCO$^-$ can form through a sequential dissociation process. In the first step, the C$-$O bond dissociates, leading to the formation of CH$_3$CO$^-$ and the neutral conjugate C$_2$H$_5$O (as shown in Channel 4a). In the second step, CH$_3$CO$^-$ further dissociates to form HCCO$^-$ and H$_2$. The process can be represented as follows:
\begin{align*}
(\mbox{C}\mbox{H}_{3}\mbox{COO}\mbox{C}_{2}\mbox{H}_{5})^{*-} 
& \rightarrow \mbox{C}\mbox{H}_3\mbox{CO}^{-} + \mbox{C}_{2}\mbox{H}_{5}\mbox{O}  \\
& \rightarrow \mbox{H}\mbox{C}\mbox{CO}^{-} + \mbox{H}_2 + \mbox{C}_{2}\mbox{H}_{5}\mbox{O} \tag{6a}
\end{align*}
The computed threshold for the concerted three-body dissociation processes is 3.85 eV, indicating that the 3.3 eV \emph{shape resonance} can not proceed through this channel. Alternatively, for the intermediate step, the computed threshold energy is 3.90 eV (channel 4a). As a result, the formation of HCCO$^-$ from the 3.3 eV \emph{shape resonance} can be ruled out through the sequential dissociation channel (6a) as well. This suggests that this channel (both sequential and concerted) is only relevant for the 6-12 eV \emph{Feshbach resonances}.\\

Another possibility arises from the C$_2$H$_5$O$^-$ fragment. During the C$-$O dissociation near the keto site, C$_2$H$_5$O$^-$ is also a potential fragment (as shown in Channel 3a). This C$_2$H$_5$O$^-$ ion can sequentially dissociate, producing the HCCO$^-$ ions. In this fragmentation process, four C$-$H bonds can dissociate simultaneously, accompanied by the formation of two H$-$H bonds. The process can be represented as follows:
\begin{align*}
(\mbox{C}\mbox{H}_{3}\mbox{COO}\mbox{C}_{2}\mbox{H}_{5})^{*-} 
& \rightarrow \mbox{C}\mbox{H}_3\mbox{CO} + \mbox{C}_{2}\mbox{H}_{5}\mbox{O}^{-}  \\
& \rightarrow \mbox{C}\mbox{H}_3\mbox{CO} + \mbox{H}\mbox{C}\mbox{CO}^{-} + 2\mbox{H}_2 \tag{6b}
\end{align*}
As discussed earlier, this dissociation can occur either as a concerted process through structural rearrangements in the TNI or as a sequential dissociation process. The computed threshold for this dissociation channel is 3.85 eV. Consequently, we can exclude channel (6b) for the production of HCCO$^-$ associated with the 3.3 eV resonance.\\

This observation indicates that, for the 3.3 eV \emph{shape resonance}, the formation of the HCOO$^-$ ion is not possible when the dissociation occurs near the keto side, producing molecular H$_2$ as a by-product. Both channels (6a and 6b) are relevant only to the 6-12 eV \emph{Feshbach resonance}.\\
\subsubsection{C$-$O dissociation near the ethyl site}
In this dissociation channel, water (H$_2$O) forms as a by-product. This dissociation involves a concerted three-body process through structural rearrangements in the TNI. In this process, two C$-$H bonds dissociate from the CH$_3$COO site, one C$-$O bond dissociates, and two O$-$H bonds form, resulting in the production of H$_2$O (Channel 6c). The formation of HCCO$^-$ from the acetate (CH$_3$COO) site has been observed in previous studies \cite{Sailer_acetic_acid,Chakraborty_AceticAcid}. In a recent report on DEA to acetic acid, Chakraborty \emph{et al.} \cite{Chakraborty_AceticAcid} identified a strong signal for the HCCO$^-$ ions around 2 eV and 10 eV regions, which exactly matches the structure of HCCO$^-$ ions observed in the present study. The authors attribute the formation of HCCO$^-$ to structural rearrangements in the acetic acid TNI, where H$_2$O is formed as a by-product. By comparing with the previous report, we assign Channel (6c) as the most probable fragmentation channel for the HCCO$^-$ formation. We discard the sequential dissociation process through the formation of the CH$_3$COO$^-$ due to its extremely stable configuration. The computed threshold energy for this channel is 2.73 eV.
\begin{align*}
(\mbox{C}\mbox{H}_{3}\mbox{COO}\mbox{C}_{2}\mbox{H}_{5})^{*-} 
& \rightarrow \mbox{H}\mbox{C}\mbox{CO}^{-} + \mbox{H}_{2}\mbox{O} + \mbox{C}_{2}\mbox{H}_{5} \tag{6c}
\end{align*}
This observation indicates that, for the 3.3 eV \emph{shape resonance}, this is the sole dissociation channel capable of producing HCCO$^-$. This dissociation is also highly probable for the 6-12 eV resonance as well.
\section{Conclusions}
The investigation of DEA to ethyl acetate within the 1 to 13 eV energy range using a QMS revealed a \emph{shape resonance} around 3 eV, accompanied by multiple \emph{core-excited Feshbach} resonances in the 5 to 12 eV energy region. Fragment anions with six different masses were observed, and the production of [(M$-$H)$^-$], CH$_3^-$, CH$_3$CO$^-$, C$_2$H$_5$O$^-$ and CH$_3$COO$^-$ ions was justified by single bond breakage. However, the formation of HCCO$^-$ involved a complex dissociation mechanism. Our experimental findings, combined with theoretical calculations, provide insight into the potential dissociation mechanisms leading to HCCO$^-$ formation. It is suggested that three-body or higher-order dissociations may be involved in all these dissociation channels.

\section{Acknowledgements}
D.C., S.P., and I.C. are supported by the U.S. Department of Energy, Office of Science, Office of Basic Energy Sciences under Award No. DE-FC02-04ER15533 (NDRL No. 5420) for the work performed at Notre Dame Radiation Laboratory. A.P. deeply appreciates the ``Council of Scientific and Industrial Research (CSIR)'' for the financial assistance.

\bibliography{ethyl_acetate.bib}

\begin{thebibliography}{29}%
\makeatletter
\providecommand \@ifxundefined [1]{%
 \@ifx{#1\undefined}
}%
\providecommand \@ifnum [1]{%
 \ifnum #1\expandafter \@firstoftwo
 \else \expandafter \@secondoftwo
 \fi
}%
\providecommand \@ifx [1]{%
 \ifx #1\expandafter \@firstoftwo
 \else \expandafter \@secondoftwo
 \fi
}%
\providecommand \natexlab [1]{#1}%
\providecommand \enquote  [1]{``#1''}%
\providecommand \bibnamefont  [1]{#1}%
\providecommand \bibfnamefont [1]{#1}%
\providecommand \citenamefont [1]{#1}%
\providecommand \href@noop [0]{\@secondoftwo}%
\providecommand \href [0]{\begingroup \@sanitize@url \@href}%
\providecommand \@href[1]{\@@startlink{#1}\@@href}%
\providecommand \@@href[1]{\endgroup#1\@@endlink}%
\providecommand \@sanitize@url [0]{\catcode `\\12\catcode `\$12\catcode
  `\&12\catcode `\#12\catcode `\^12\catcode `\_12\catcode `\%12\relax}%
\providecommand \@@startlink[1]{}%
\providecommand \@@endlink[0]{}%
\providecommand \url  [0]{\begingroup\@sanitize@url \@url }%
\providecommand \@url [1]{\endgroup\@href {#1}{\urlprefix }}%
\providecommand \urlprefix  [0]{URL }%
\providecommand \Eprint [0]{\href }%
\providecommand \doibase [0]{http://dx.doi.org/}%
\providecommand \selectlanguage [0]{\@gobble}%
\providecommand \bibinfo  [0]{\@secondoftwo}%
\providecommand \bibfield  [0]{\@secondoftwo}%
\providecommand \translation [1]{[#1]}%
\providecommand \BibitemOpen [0]{}%
\providecommand \bibitemStop [0]{}%
\providecommand \bibitemNoStop [0]{.\EOS\space}%
\providecommand \EOS [0]{\spacefactor3000\relax}%
\providecommand \BibitemShut  [1]{\csname bibitem#1\endcsname}%
\let\auto@bib@innerbib\@empty
\bibitem [{\citenamefont {Lu}\ and\ \citenamefont {Sanche}(2001)}]{Lu2001}%
  \BibitemOpen
  \bibfield  {author} {\bibinfo {author} {\bibfnamefont {Q.~B.}\ \bibnamefont
  {Lu}}\ and\ \bibinfo {author} {\bibfnamefont {L.}~\bibnamefont {Sanche}},\
  }\href {\doibase 10.1103/PhysRevLett.87.078501} {\bibfield  {journal}
  {\bibinfo  {journal} {Phys. Rev. Lett.}\ }\textbf {\bibinfo {volume} {87}},\
  \bibinfo {pages} {078501} (\bibinfo {year} {2001})}\BibitemShut {NoStop}%
\bibitem [{\citenamefont {Boudaiffa}\ \emph {et~al.}(2000)\citenamefont
  {Boudaiffa}, \citenamefont {Cloutier}, \citenamefont {Hunting}, \citenamefont
  {Huels},\ and\ \citenamefont {Sanche}}]{Boudaiffa}%
  \BibitemOpen
  \bibfield  {author} {\bibinfo {author} {\bibfnamefont {B.}~\bibnamefont
  {Boudaiffa}}, \bibinfo {author} {\bibfnamefont {P.}~\bibnamefont {Cloutier}},
  \bibinfo {author} {\bibfnamefont {D.}~\bibnamefont {Hunting}}, \bibinfo
  {author} {\bibfnamefont {M.~A.}\ \bibnamefont {Huels}}, \ and\ \bibinfo
  {author} {\bibfnamefont {L.}~\bibnamefont {Sanche}},\ }\href {\doibase
  10.1126/science.287.5458.1658} {\bibfield  {journal} {\bibinfo  {journal}
  {Sci.}\ }\textbf {\bibinfo {volume} {287}},\ \bibinfo {pages} {1658}
  (\bibinfo {year} {2000})}\BibitemShut {NoStop}%
\bibitem [{\citenamefont {Illenberger}\ and\ \citenamefont
  {Momigny}(2014)}]{illenberger2014gaseous}%
  \BibitemOpen
  \bibfield  {author} {\bibinfo {author} {\bibfnamefont {E.}~\bibnamefont
  {Illenberger}}\ and\ \bibinfo {author} {\bibfnamefont {J.}~\bibnamefont
  {Momigny}},\ }\href {https://books.google.com.cu/books?id=MqQmswEACAAJ}
  {\emph {\bibinfo {title} {Gaseous Molecular Ions: An Introduction to
  Elementary Processes Induced by Ionization}}},\ Topics in Physical Chemistry\
  (\bibinfo  {publisher} {Steinkopff},\ \bibinfo {year} {2014})\BibitemShut
  {NoStop}%
\bibitem [{\citenamefont {Ptasinska}\ \emph {et~al.}(2022)\citenamefont
  {Ptasinska}, \citenamefont {Varella}, \citenamefont {Khakoo}, \citenamefont
  {Slaughter},\ and\ \citenamefont {Denifl}}]{ptasinska_epjd}%
  \BibitemOpen
  \bibfield  {author} {\bibinfo {author} {\bibfnamefont {S.}~\bibnamefont
  {Ptasinska}}, \bibinfo {author} {\bibfnamefont {M.~T. d.~N.}\ \bibnamefont
  {Varella}}, \bibinfo {author} {\bibfnamefont {M.~A.}\ \bibnamefont {Khakoo}},
  \bibinfo {author} {\bibfnamefont {D.~S.}\ \bibnamefont {Slaughter}}, \ and\
  \bibinfo {author} {\bibfnamefont {S.}~\bibnamefont {Denifl}},\ }\href
  {\doibase 10.1140/epjd/s10053-022-00482-8} {\bibfield  {journal} {\bibinfo
  {journal} {Eur. Phys. J. D}\ }\textbf {\bibinfo {volume} {76}},\ \bibinfo
  {pages} {179} (\bibinfo {year} {2022})}\BibitemShut {NoStop}%
\bibitem [{\citenamefont {Riemenschneider}\ and\ \citenamefont
  {Bolt}(2005)}]{Riemenschneider2005}%
  \BibitemOpen
  \bibfield  {author} {\bibinfo {author} {\bibfnamefont {W.}~\bibnamefont
  {Riemenschneider}}\ and\ \bibinfo {author} {\bibfnamefont {H.~M.}\
  \bibnamefont {Bolt}},\ }\href
  {https://onlinelibrary.wiley.com/doi/abs/10.1002/14356007.a09_565.pub2}
  {\emph {\bibinfo {title} {Ullmann's Encyclopedia of Industrial Chemistry}}},\
  Esters, Organic\ (\bibinfo  {publisher} {John Wiley $\&$ Sons, Ltd},\
  \bibinfo {year} {2005})\BibitemShut {NoStop}%
\bibitem [{\citenamefont {Buxton}\ \emph {et~al.}(2001)\citenamefont {Buxton},
  \citenamefont {Wang},\ and\ \citenamefont {Salmon}}]{buxton2001rate}%
  \BibitemOpen
  \bibfield  {author} {\bibinfo {author} {\bibfnamefont {G.~V.}\ \bibnamefont
  {Buxton}}, \bibinfo {author} {\bibfnamefont {J.}~\bibnamefont {Wang}}, \ and\
  \bibinfo {author} {\bibfnamefont {G.~A.}\ \bibnamefont {Salmon}},\
  }\href@noop {} {\bibfield  {journal} {\bibinfo  {journal} {Phys. Chem. Chem.
  Phys.}\ }\textbf {\bibinfo {volume} {3}},\ \bibinfo {pages} {2618} (\bibinfo
  {year} {2001})}\BibitemShut {NoStop}%
\bibitem [{\citenamefont {Schneider}\ \emph {et~al.}(2001)\citenamefont
  {Schneider}, \citenamefont {Gohla}, \citenamefont {Schreiber}, \citenamefont
  {Kaden}, \citenamefont {Schönrock}, \citenamefont {Schmidt-Lewerkühne},
  \citenamefont {Kuschel}, \citenamefont {Petsitis}, \citenamefont {Pape},
  \citenamefont {Ippen},\ and\ \citenamefont {Diembeck}}]{Schneider2001}%
  \BibitemOpen
  \bibfield  {author} {\bibinfo {author} {\bibfnamefont {G.}~\bibnamefont
  {Schneider}}, \bibinfo {author} {\bibfnamefont {S.}~\bibnamefont {Gohla}},
  \bibinfo {author} {\bibfnamefont {J.}~\bibnamefont {Schreiber}}, \bibinfo
  {author} {\bibfnamefont {W.}~\bibnamefont {Kaden}}, \bibinfo {author}
  {\bibfnamefont {U.}~\bibnamefont {Schönrock}}, \bibinfo {author}
  {\bibfnamefont {H.}~\bibnamefont {Schmidt-Lewerkühne}}, \bibinfo {author}
  {\bibfnamefont {A.}~\bibnamefont {Kuschel}}, \bibinfo {author} {\bibfnamefont
  {X.}~\bibnamefont {Petsitis}}, \bibinfo {author} {\bibfnamefont
  {W.}~\bibnamefont {Pape}}, \bibinfo {author} {\bibfnamefont {H.}~\bibnamefont
  {Ippen}}, \ and\ \bibinfo {author} {\bibfnamefont {W.}~\bibnamefont
  {Diembeck}},\ }\href
  {https://onlinelibrary.wiley.com/doi/abs/10.1002/14356007.a24_219} {\emph
  {\bibinfo {title} {Skin Cosmetics}}}\ (\bibinfo  {publisher} {John Wiley $\&$
  Sons, Ltd},\ \bibinfo {year} {2001})\BibitemShut {NoStop}%
\bibitem [{\citenamefont {Ramalakshmi}\ and\ \citenamefont
  {Raghavan}(1999)}]{Ramalakshmi1999}%
  \BibitemOpen
  \bibfield  {author} {\bibinfo {author} {\bibfnamefont {K.}~\bibnamefont
  {Ramalakshmi}}\ and\ \bibinfo {author} {\bibfnamefont {B.}~\bibnamefont
  {Raghavan}},\ }\href {\doibase 10.1080/10408699991279231} {\bibfield
  {journal} {\bibinfo  {journal} {Crit. Rev. Food. Sci. Nutr.}\ }\textbf
  {\bibinfo {volume} {39}},\ \bibinfo {pages} {441} (\bibinfo {year}
  {1999})}\BibitemShut {NoStop}%
\bibitem [{\citenamefont {Tan}\ \emph {et~al.}(2003)\citenamefont {Tan},
  \citenamefont {Wu}, \citenamefont {Wei},\ and\ \citenamefont
  {Yoshikai}}]{tan2003copper}%
  \BibitemOpen
  \bibfield  {author} {\bibinfo {author} {\bibfnamefont {W.~W.}\ \bibnamefont
  {Tan}}, \bibinfo {author} {\bibfnamefont {B.}~\bibnamefont {Wu}}, \bibinfo
  {author} {\bibfnamefont {Y.}~\bibnamefont {Wei}}, \ and\ \bibinfo {author}
  {\bibfnamefont {N.}~\bibnamefont {Yoshikai}},\ }\href@noop {} {\bibfield
  {journal} {\bibinfo  {journal} {Org. Synth.}\ }\textbf {\bibinfo {volume}
  {95}},\ \bibinfo {pages} {1} (\bibinfo {year} {2003})}\BibitemShut {NoStop}%
\bibitem [{\citenamefont {Helmig}\ \emph {et~al.}(1989)\citenamefont {Helmig},
  \citenamefont {M{\"u}ller},\ and\ \citenamefont
  {Klein}}]{helmig1989volatile}%
  \BibitemOpen
  \bibfield  {author} {\bibinfo {author} {\bibfnamefont {D.}~\bibnamefont
  {Helmig}}, \bibinfo {author} {\bibfnamefont {J.}~\bibnamefont {M{\"u}ller}},
  \ and\ \bibinfo {author} {\bibfnamefont {W.}~\bibnamefont {Klein}},\
  }\href@noop {} {\bibfield  {journal} {\bibinfo  {journal} {Chemosphere}\
  }\textbf {\bibinfo {volume} {19}},\ \bibinfo {pages} {1399} (\bibinfo {year}
  {1989})}\BibitemShut {NoStop}%
\bibitem [{\citenamefont {Pariat}\ and\ \citenamefont
  {Allan}(1991)}]{PARIAT1991181}%
  \BibitemOpen
  \bibfield  {author} {\bibinfo {author} {\bibfnamefont {Y.}~\bibnamefont
  {Pariat}}\ and\ \bibinfo {author} {\bibfnamefont {M.}~\bibnamefont {Allan}},\
  }\href {\doibase https://doi.org/10.1016/0168-1176(91)80088-5} {\bibfield
  {journal} {\bibinfo  {journal} {Int. J. Mass Spectrom. Ion Processes}\
  }\textbf {\bibinfo {volume} {103}},\ \bibinfo {pages} {181} (\bibinfo {year}
  {1991})}\BibitemShut {NoStop}%
\bibitem [{\citenamefont {{Feketeov\'a, L.}}\ \emph {et~al.}(2018)\citenamefont
  {{Feketeov\'a, L.}}, \citenamefont {{Pelc, A.}}, \citenamefont {{Ribar, A.}},
  \citenamefont {{Huber, S. E.}},\ and\ \citenamefont {{Denifl,
  S.}}}]{Feketova2018}%
  \BibitemOpen
  \bibfield  {author} {\bibinfo {author} {\bibnamefont {{Feketeov\'a, L.}}},
  \bibinfo {author} {\bibnamefont {{Pelc, A.}}}, \bibinfo {author}
  {\bibnamefont {{Ribar, A.}}}, \bibinfo {author} {\bibnamefont {{Huber, S.
  E.}}}, \ and\ \bibinfo {author} {\bibnamefont {{Denifl, S.}}},\ }\href
  {\doibase 10.1051/0004-6361/201732293} {\bibfield  {journal} {\bibinfo
  {journal} {A\&A}\ }\textbf {\bibinfo {volume} {617}},\ \bibinfo {pages}
  {A102} (\bibinfo {year} {2018})}\BibitemShut {NoStop}%
\bibitem [{\citenamefont {K\"onig}\ \emph {et~al.}(2006)\citenamefont
  {K\"onig}, \citenamefont {Kopyra}, \citenamefont {Bald},\ and\ \citenamefont
  {Illenberger}}]{Illenberger2006}%
  \BibitemOpen
  \bibfield  {author} {\bibinfo {author} {\bibfnamefont {C.}~\bibnamefont
  {K\"onig}}, \bibinfo {author} {\bibfnamefont {J.}~\bibnamefont {Kopyra}},
  \bibinfo {author} {\bibfnamefont {I.}~\bibnamefont {Bald}}, \ and\ \bibinfo
  {author} {\bibfnamefont {E.}~\bibnamefont {Illenberger}},\ }\href {\doibase
  10.1103/PhysRevLett.97.018105} {\bibfield  {journal} {\bibinfo  {journal}
  {Phys. Rev. Lett.}\ }\textbf {\bibinfo {volume} {97}},\ \bibinfo {pages}
  {018105} (\bibinfo {year} {2006})}\BibitemShut {NoStop}%
\bibitem [{\citenamefont {Chakraborty}\ \emph {et~al.}(2020)\citenamefont
  {Chakraborty}, \citenamefont {Eckermann}, \citenamefont {Carmichael},\ and\
  \citenamefont {Ptasińska}}]{Chakraborty_NMF}%
  \BibitemOpen
  \bibfield  {author} {\bibinfo {author} {\bibfnamefont {D.}~\bibnamefont
  {Chakraborty}}, \bibinfo {author} {\bibfnamefont {L.}~\bibnamefont
  {Eckermann}}, \bibinfo {author} {\bibfnamefont {I.}~\bibnamefont
  {Carmichael}}, \ and\ \bibinfo {author} {\bibfnamefont {S.}~\bibnamefont
  {Ptasińska}},\ }\href {\doibase 10.1063/5.0029614} {\bibfield  {journal}
  {\bibinfo  {journal} {J Chem. Phys.}\ }\textbf {\bibinfo {volume} {153}},\
  \bibinfo {pages} {224306} (\bibinfo {year} {2020})}\BibitemShut {NoStop}%
\bibitem [{\citenamefont {Rapp}\ and\ \citenamefont {Briglia}(1965)}]{rapp}%
  \BibitemOpen
  \bibfield  {author} {\bibinfo {author} {\bibfnamefont {D.}~\bibnamefont
  {Rapp}}\ and\ \bibinfo {author} {\bibfnamefont {D.~D.}\ \bibnamefont
  {Briglia}},\ }\href {\doibase 10.1063/1.1696958} {\bibfield  {journal}
  {\bibinfo  {journal} {J. Chem. Phys.}\ }\textbf {\bibinfo {volume} {43}},\
  \bibinfo {pages} {1480} (\bibinfo {year} {1965})}\BibitemShut {NoStop}%
\bibitem [{\citenamefont {Frisch}\ \emph {et~al.}(2016)\citenamefont {Frisch},
  \citenamefont {Trucks}, \citenamefont {Schlegel}, \citenamefont {Scuseria},
  \citenamefont {Robb}, \citenamefont {Cheeseman}, \citenamefont {Scalmani},
  \citenamefont {Barone}, \citenamefont {Petersson}, \citenamefont {Nakatsuji},
  \citenamefont {Li}, \citenamefont {Caricato}, \citenamefont {Marenich},
  \citenamefont {Bloino}, \citenamefont {Janesko}, \citenamefont {Gomperts},
  \citenamefont {Mennucci}, \citenamefont {Hratchian}, \citenamefont {Ortiz},
  \citenamefont {Izmaylov}, \citenamefont {Sonnenberg}, \citenamefont
  {Williams-Young}, \citenamefont {Ding}, \citenamefont {Lipparini},
  \citenamefont {Egidi}, \citenamefont {Goings}, \citenamefont {Peng},
  \citenamefont {Petrone}, \citenamefont {Henderson}, \citenamefont
  {Ranasinghe}, \citenamefont {Zakrzewski}, \citenamefont {Gao}, \citenamefont
  {Rega}, \citenamefont {Zheng}, \citenamefont {Liang}, \citenamefont {Hada},
  \citenamefont {Ehara}, \citenamefont {Toyota}, \citenamefont {Fukuda},
  \citenamefont {Hasegawa}, \citenamefont {Ishida}, \citenamefont {Nakajima},
  \citenamefont {Honda}, \citenamefont {Kitao}, \citenamefont {Nakai},
  \citenamefont {Vreven}, \citenamefont {Throssell}, \citenamefont
  {Montgomery}, \citenamefont {Peralta}, \citenamefont {Ogliaro}, \citenamefont
  {Bearpark}, \citenamefont {Heyd}, \citenamefont {Brothers}, \citenamefont
  {Kudin}, \citenamefont {Staroverov}, \citenamefont {Keith}, \citenamefont
  {Kobayashi}, \citenamefont {Normand}, \citenamefont {Raghavachari},
  \citenamefont {Rendell}, \citenamefont {Burant}, \citenamefont {Iyengar},
  \citenamefont {Tomasi}, \citenamefont {Cossi}, \citenamefont {Millam},
  \citenamefont {Klene}, \citenamefont {Adamo}, \citenamefont {Cammi},
  \citenamefont {Ochterski}, \citenamefont {Martin}, \citenamefont {Morokuma},
  \citenamefont {Farkas}, \citenamefont {Foresman},\ and\ \citenamefont
  {Fox}}]{g16}%
  \BibitemOpen
  \bibfield  {author} {\bibinfo {author} {\bibfnamefont {M.~J.}\ \bibnamefont
  {Frisch}}, \bibinfo {author} {\bibfnamefont {G.~W.}\ \bibnamefont {Trucks}},
  \bibinfo {author} {\bibfnamefont {H.~B.}\ \bibnamefont {Schlegel}}, \bibinfo
  {author} {\bibfnamefont {G.~E.}\ \bibnamefont {Scuseria}}, \bibinfo {author}
  {\bibfnamefont {M.~A.}\ \bibnamefont {Robb}}, \bibinfo {author}
  {\bibfnamefont {J.~R.}\ \bibnamefont {Cheeseman}}, \bibinfo {author}
  {\bibfnamefont {G.}~\bibnamefont {Scalmani}}, \bibinfo {author}
  {\bibfnamefont {V.}~\bibnamefont {Barone}}, \bibinfo {author} {\bibfnamefont
  {G.~A.}\ \bibnamefont {Petersson}}, \bibinfo {author} {\bibfnamefont
  {H.}~\bibnamefont {Nakatsuji}}, \bibinfo {author} {\bibfnamefont
  {X.}~\bibnamefont {Li}}, \bibinfo {author} {\bibfnamefont {M.}~\bibnamefont
  {Caricato}}, \bibinfo {author} {\bibfnamefont {A.~V.}\ \bibnamefont
  {Marenich}}, \bibinfo {author} {\bibfnamefont {J.}~\bibnamefont {Bloino}},
  \bibinfo {author} {\bibfnamefont {B.~G.}\ \bibnamefont {Janesko}}, \bibinfo
  {author} {\bibfnamefont {R.}~\bibnamefont {Gomperts}}, \bibinfo {author}
  {\bibfnamefont {B.}~\bibnamefont {Mennucci}}, \bibinfo {author}
  {\bibfnamefont {H.~P.}\ \bibnamefont {Hratchian}}, \bibinfo {author}
  {\bibfnamefont {J.~V.}\ \bibnamefont {Ortiz}}, \bibinfo {author}
  {\bibfnamefont {A.~F.}\ \bibnamefont {Izmaylov}}, \bibinfo {author}
  {\bibfnamefont {J.~L.}\ \bibnamefont {Sonnenberg}}, \bibinfo {author}
  {\bibfnamefont {D.}~\bibnamefont {Williams-Young}}, \bibinfo {author}
  {\bibfnamefont {F.}~\bibnamefont {Ding}}, \bibinfo {author} {\bibfnamefont
  {F.}~\bibnamefont {Lipparini}}, \bibinfo {author} {\bibfnamefont
  {F.}~\bibnamefont {Egidi}}, \bibinfo {author} {\bibfnamefont
  {J.}~\bibnamefont {Goings}}, \bibinfo {author} {\bibfnamefont
  {B.}~\bibnamefont {Peng}}, \bibinfo {author} {\bibfnamefont {A.}~\bibnamefont
  {Petrone}}, \bibinfo {author} {\bibfnamefont {T.}~\bibnamefont {Henderson}},
  \bibinfo {author} {\bibfnamefont {D.}~\bibnamefont {Ranasinghe}}, \bibinfo
  {author} {\bibfnamefont {V.~G.}\ \bibnamefont {Zakrzewski}}, \bibinfo
  {author} {\bibfnamefont {J.}~\bibnamefont {Gao}}, \bibinfo {author}
  {\bibfnamefont {N.}~\bibnamefont {Rega}}, \bibinfo {author} {\bibfnamefont
  {G.}~\bibnamefont {Zheng}}, \bibinfo {author} {\bibfnamefont
  {W.}~\bibnamefont {Liang}}, \bibinfo {author} {\bibfnamefont
  {M.}~\bibnamefont {Hada}}, \bibinfo {author} {\bibfnamefont {M.}~\bibnamefont
  {Ehara}}, \bibinfo {author} {\bibfnamefont {K.}~\bibnamefont {Toyota}},
  \bibinfo {author} {\bibfnamefont {R.}~\bibnamefont {Fukuda}}, \bibinfo
  {author} {\bibfnamefont {J.}~\bibnamefont {Hasegawa}}, \bibinfo {author}
  {\bibfnamefont {M.}~\bibnamefont {Ishida}}, \bibinfo {author} {\bibfnamefont
  {T.}~\bibnamefont {Nakajima}}, \bibinfo {author} {\bibfnamefont
  {Y.}~\bibnamefont {Honda}}, \bibinfo {author} {\bibfnamefont
  {O.}~\bibnamefont {Kitao}}, \bibinfo {author} {\bibfnamefont
  {H.}~\bibnamefont {Nakai}}, \bibinfo {author} {\bibfnamefont
  {T.}~\bibnamefont {Vreven}}, \bibinfo {author} {\bibfnamefont
  {K.}~\bibnamefont {Throssell}}, \bibinfo {author} {\bibfnamefont {J.~A.}\
  \bibnamefont {Montgomery}, \bibfnamefont {{Jr.}}}, \bibinfo {author}
  {\bibfnamefont {J.~E.}\ \bibnamefont {Peralta}}, \bibinfo {author}
  {\bibfnamefont {F.}~\bibnamefont {Ogliaro}}, \bibinfo {author} {\bibfnamefont
  {M.~J.}\ \bibnamefont {Bearpark}}, \bibinfo {author} {\bibfnamefont {J.~J.}\
  \bibnamefont {Heyd}}, \bibinfo {author} {\bibfnamefont {E.~N.}\ \bibnamefont
  {Brothers}}, \bibinfo {author} {\bibfnamefont {K.~N.}\ \bibnamefont {Kudin}},
  \bibinfo {author} {\bibfnamefont {V.~N.}\ \bibnamefont {Staroverov}},
  \bibinfo {author} {\bibfnamefont {T.~A.}\ \bibnamefont {Keith}}, \bibinfo
  {author} {\bibfnamefont {R.}~\bibnamefont {Kobayashi}}, \bibinfo {author}
  {\bibfnamefont {J.}~\bibnamefont {Normand}}, \bibinfo {author} {\bibfnamefont
  {K.}~\bibnamefont {Raghavachari}}, \bibinfo {author} {\bibfnamefont {A.~P.}\
  \bibnamefont {Rendell}}, \bibinfo {author} {\bibfnamefont {J.~C.}\
  \bibnamefont {Burant}}, \bibinfo {author} {\bibfnamefont {S.~S.}\
  \bibnamefont {Iyengar}}, \bibinfo {author} {\bibfnamefont {J.}~\bibnamefont
  {Tomasi}}, \bibinfo {author} {\bibfnamefont {M.}~\bibnamefont {Cossi}},
  \bibinfo {author} {\bibfnamefont {J.~M.}\ \bibnamefont {Millam}}, \bibinfo
  {author} {\bibfnamefont {M.}~\bibnamefont {Klene}}, \bibinfo {author}
  {\bibfnamefont {C.}~\bibnamefont {Adamo}}, \bibinfo {author} {\bibfnamefont
  {R.}~\bibnamefont {Cammi}}, \bibinfo {author} {\bibfnamefont {J.~W.}\
  \bibnamefont {Ochterski}}, \bibinfo {author} {\bibfnamefont {R.~L.}\
  \bibnamefont {Martin}}, \bibinfo {author} {\bibfnamefont {K.}~\bibnamefont
  {Morokuma}}, \bibinfo {author} {\bibfnamefont {O.}~\bibnamefont {Farkas}},
  \bibinfo {author} {\bibfnamefont {J.~B.}\ \bibnamefont {Foresman}}, \ and\
  \bibinfo {author} {\bibfnamefont {D.~J.}\ \bibnamefont {Fox}},\ }\href@noop
  {} {\enquote {\bibinfo {title} {Gaussian˜16 {R}evision {C}.01},}\ }
  (\bibinfo {year} {2016}),\ \bibinfo {note} {gaussian Inc. Wallingford
  CT}\BibitemShut {NoStop}%
\bibitem [{\citenamefont {Martin}\ and\ \citenamefont
  {de~Oliveira}(1999)}]{W1_method}%
  \BibitemOpen
  \bibfield  {author} {\bibinfo {author} {\bibfnamefont {J.~M.~L.}\
  \bibnamefont {Martin}}\ and\ \bibinfo {author} {\bibfnamefont
  {G.}~\bibnamefont {de~Oliveira}},\ }\href {\doibase 10.1063/1.479454}
  {\bibfield  {journal} {\bibinfo  {journal} {J. Chem. Phys.}\ }\textbf
  {\bibinfo {volume} {111}},\ \bibinfo {pages} {1843} (\bibinfo {year}
  {1999})}\BibitemShut {NoStop}%
\bibitem [{\citenamefont {Becke}(1993)}]{B3LYP}%
  \BibitemOpen
  \bibfield  {author} {\bibinfo {author} {\bibfnamefont {A.~D.}\ \bibnamefont
  {Becke}},\ }\href {\doibase 10.1063/1.464913} {\bibfield  {journal} {\bibinfo
   {journal} {J. Chem. Phys.}\ }\textbf {\bibinfo {volume} {98}},\ \bibinfo
  {pages} {5648} (\bibinfo {year} {1993})}\BibitemShut {NoStop}%
\bibitem [{\citenamefont {Dunning}(1989)}]{Dunning1989}%
  \BibitemOpen
  \bibfield  {author} {\bibinfo {author} {\bibfnamefont {T.~H.}\ \bibnamefont
  {Dunning}},\ }\href {\doibase 10.1063/1.456153} {\bibfield  {journal}
  {\bibinfo  {journal} {J. Chem. Phys.}\ }\textbf {\bibinfo {volume} {90}},\
  \bibinfo {pages} {1007} (\bibinfo {year} {1989})}\BibitemShut {NoStop}%
\bibitem [{\citenamefont {Barnes}\ \emph {et~al.}(2009)\citenamefont {Barnes},
  \citenamefont {Petersson}, \citenamefont {Montgomery}, \citenamefont
  {Frisch},\ and\ \citenamefont {Martin}}]{Barnes_2009}%
  \BibitemOpen
  \bibfield  {author} {\bibinfo {author} {\bibfnamefont {E.~C.}\ \bibnamefont
  {Barnes}}, \bibinfo {author} {\bibfnamefont {G.~A.}\ \bibnamefont
  {Petersson}}, \bibinfo {author} {\bibfnamefont {J.~A.~J.}\ \bibnamefont
  {Montgomery}}, \bibinfo {author} {\bibfnamefont {M.~J.}\ \bibnamefont
  {Frisch}}, \ and\ \bibinfo {author} {\bibfnamefont {J.~M.~L.}\ \bibnamefont
  {Martin}},\ }\href {\doibase 10.1021/ct900260g} {\bibfield  {journal}
  {\bibinfo  {journal} {J. Chem. Theory Comput.}\ }\textbf {\bibinfo {volume}
  {5}},\ \bibinfo {pages} {2687} (\bibinfo {year} {2009})},\ \bibinfo {note}
  {pMID: 26631782}\BibitemShut {NoStop}%
\bibitem [{\citenamefont {Śmialek}\ \emph {et~al.}(2016)\citenamefont
  {Śmialek}, \citenamefont {Łabuda}, \citenamefont {Guthmuller},
  \citenamefont {Hubin-Franskin}, \citenamefont {Delwiche}, \citenamefont
  {Hoffmann}, \citenamefont {Jones}, \citenamefont {Mason},\ and\ \citenamefont
  {Limão-Vieira}}]{Paulo2016}%
  \BibitemOpen
  \bibfield  {author} {\bibinfo {author} {\bibfnamefont {M.~A.}\ \bibnamefont
  {Śmialek}}, \bibinfo {author} {\bibfnamefont {M.}~\bibnamefont {Łabuda}},
  \bibinfo {author} {\bibfnamefont {J.}~\bibnamefont {Guthmuller}}, \bibinfo
  {author} {\bibfnamefont {M.-J.}\ \bibnamefont {Hubin-Franskin}}, \bibinfo
  {author} {\bibfnamefont {J.}~\bibnamefont {Delwiche}}, \bibinfo {author}
  {\bibfnamefont {S.~V.}\ \bibnamefont {Hoffmann}}, \bibinfo {author}
  {\bibfnamefont {N.~C.}\ \bibnamefont {Jones}}, \bibinfo {author}
  {\bibfnamefont {N.~J.}\ \bibnamefont {Mason}}, \ and\ \bibinfo {author}
  {\bibfnamefont {P.}~\bibnamefont {Limão-Vieira}},\ }\href {\doibase
  10.1140/epjd/e2016-70239-9} {\bibfield  {journal} {\bibinfo  {journal} {Eur.
  Phys. J. D.}\ }\textbf {\bibinfo {volume} {70}},\ \bibinfo {pages} {138}
  (\bibinfo {year} {2016})}\BibitemShut {NoStop}%
\bibitem [{\citenamefont {Chakraborty}\ \emph {et~al.}(2016)\citenamefont
  {Chakraborty}, \citenamefont {Nag},\ and\ \citenamefont
  {Nandi}}]{ChakrabortyIPD}%
  \BibitemOpen
  \bibfield  {author} {\bibinfo {author} {\bibfnamefont {D.}~\bibnamefont
  {Chakraborty}}, \bibinfo {author} {\bibfnamefont {P.}~\bibnamefont {Nag}}, \
  and\ \bibinfo {author} {\bibfnamefont {D.}~\bibnamefont {Nandi}},\ }\href
  {\doibase 10.1039/C6CP05854J} {\bibfield  {journal} {\bibinfo  {journal}
  {Phys. Chem. Chem. Phys.}\ }\textbf {\bibinfo {volume} {18}},\ \bibinfo
  {pages} {32973} (\bibinfo {year} {2016})}\BibitemShut {NoStop}%
\bibitem [{\citenamefont {Chakraborty}\ and\ \citenamefont
  {Paul}(2023)}]{Chakraborty_2023}%
  \BibitemOpen
  \bibfield  {author} {\bibinfo {author} {\bibfnamefont {D.}~\bibnamefont
  {Chakraborty}}\ and\ \bibinfo {author} {\bibfnamefont {A.}~\bibnamefont
  {Paul}},\ }\href {\doibase 10.1088/1361-6455/ace40d} {\bibfield  {journal}
  {\bibinfo  {journal} {J Phys. B: At. Mol. Opt. Phys.}\ }\textbf {\bibinfo
  {volume} {56}},\ \bibinfo {pages} {142001} (\bibinfo {year}
  {2023})}\BibitemShut {NoStop}%
\bibitem [{\citenamefont {Chakraborty}\ and\ \citenamefont
  {Nandi}(2020)}]{Chakraborty_CH2F2}%
  \BibitemOpen
  \bibfield  {author} {\bibinfo {author} {\bibfnamefont {D.}~\bibnamefont
  {Chakraborty}}\ and\ \bibinfo {author} {\bibfnamefont {D.}~\bibnamefont
  {Nandi}},\ }\href {\doibase 10.1103/PhysRevA.102.052801} {\bibfield
  {journal} {\bibinfo  {journal} {Phys. Rev. A}\ }\textbf {\bibinfo {volume}
  {102}},\ \bibinfo {pages} {052801} (\bibinfo {year} {2020})}\BibitemShut
  {NoStop}%
\bibitem [{\citenamefont {Ibănescu}\ \emph {et~al.}(2007)\citenamefont
  {Ibănescu}, \citenamefont {May}, \citenamefont {Monney},\ and\ \citenamefont
  {Allan}}]{Ibanescu2007}%
  \BibitemOpen
  \bibfield  {author} {\bibinfo {author} {\bibfnamefont {B.~C.}\ \bibnamefont
  {Ibănescu}}, \bibinfo {author} {\bibfnamefont {O.}~\bibnamefont {May}},
  \bibinfo {author} {\bibfnamefont {A.}~\bibnamefont {Monney}}, \ and\ \bibinfo
  {author} {\bibfnamefont {M.}~\bibnamefont {Allan}},\ }\href {\doibase
  10.1039/B704656A} {\bibfield  {journal} {\bibinfo  {journal} {Phys. Chem.
  Chem. Phys.}\ }\textbf {\bibinfo {volume} {9}},\ \bibinfo {pages} {3163}
  (\bibinfo {year} {2007})}\BibitemShut {NoStop}%
\bibitem [{\citenamefont {Paul}\ \emph {et~al.}(2023)\citenamefont {Paul},
  \citenamefont {Ghosh},\ and\ \citenamefont {Nandi}}]{Anirban_Ethanol}%
  \BibitemOpen
  \bibfield  {author} {\bibinfo {author} {\bibfnamefont {A.}~\bibnamefont
  {Paul}}, \bibinfo {author} {\bibfnamefont {S.}~\bibnamefont {Ghosh}}, \ and\
  \bibinfo {author} {\bibfnamefont {D.}~\bibnamefont {Nandi}},\ }\href
  {\doibase 10.1039/D3CP03601D} {\bibfield  {journal} {\bibinfo  {journal}
  {Phys. Chem. Chem. Phys.}\ }\textbf {\bibinfo {volume} {25}},\ \bibinfo
  {pages} {28263} (\bibinfo {year} {2023})}\BibitemShut {NoStop}%
\bibitem [{\citenamefont {Yu}\ \emph {et~al.}(2020)\citenamefont {Yu},
  \citenamefont {Wu}, \citenamefont {Zhou}, \citenamefont {Bodi},\ and\
  \citenamefont {Hemberger}}]{YU2020123}%
  \BibitemOpen
  \bibfield  {author} {\bibinfo {author} {\bibfnamefont {T.}~\bibnamefont
  {Yu}}, \bibinfo {author} {\bibfnamefont {X.}~\bibnamefont {Wu}}, \bibinfo
  {author} {\bibfnamefont {X.}~\bibnamefont {Zhou}}, \bibinfo {author}
  {\bibfnamefont {A.}~\bibnamefont {Bodi}}, \ and\ \bibinfo {author}
  {\bibfnamefont {P.}~\bibnamefont {Hemberger}},\ }\href {\doibase
  https://doi.org/10.1016/j.combustflame.2020.08.040} {\bibfield  {journal}
  {\bibinfo  {journal} {Combust. Flame.}\ }\textbf {\bibinfo {volume} {222}},\
  \bibinfo {pages} {123} (\bibinfo {year} {2020})}\BibitemShut {NoStop}%
\bibitem [{\citenamefont {Sailer}\ \emph {et~al.}(2003)\citenamefont {Sailer},
  \citenamefont {Pelc}, \citenamefont {Probst}, \citenamefont {Limtrakul},
  \citenamefont {Scheier}, \citenamefont {Illenberger},\ and\ \citenamefont
  {Märk}}]{Sailer_acetic_acid}%
  \BibitemOpen
  \bibfield  {author} {\bibinfo {author} {\bibfnamefont {W.}~\bibnamefont
  {Sailer}}, \bibinfo {author} {\bibfnamefont {A.}~\bibnamefont {Pelc}},
  \bibinfo {author} {\bibfnamefont {M.}~\bibnamefont {Probst}}, \bibinfo
  {author} {\bibfnamefont {J.}~\bibnamefont {Limtrakul}}, \bibinfo {author}
  {\bibfnamefont {P.}~\bibnamefont {Scheier}}, \bibinfo {author} {\bibfnamefont
  {E.}~\bibnamefont {Illenberger}}, \ and\ \bibinfo {author} {\bibfnamefont
  {T.~D.}\ \bibnamefont {Märk}},\ }\href {\doibase
  https://doi.org/10.1016/S0009-2614(03)01285-5} {\bibfield  {journal}
  {\bibinfo  {journal} {Chem. Phys. Lett.}\ }\textbf {\bibinfo {volume}
  {378}},\ \bibinfo {pages} {250} (\bibinfo {year} {2003})}\BibitemShut
  {NoStop}%
\bibitem [{\citenamefont {Chakraborty}\ \emph {et~al.}(2024)\citenamefont
  {Chakraborty}, \citenamefont {Kharchilava}, \citenamefont {Carmichael},\ and\
  \citenamefont {Ptasinska}}]{Chakraborty_AceticAcid}%
  \BibitemOpen
  \bibfield  {author} {\bibinfo {author} {\bibfnamefont {D.}~\bibnamefont
  {Chakraborty}}, \bibinfo {author} {\bibfnamefont {G.}~\bibnamefont
  {Kharchilava}}, \bibinfo {author} {\bibfnamefont {I.}~\bibnamefont
  {Carmichael}}, \ and\ \bibinfo {author} {\bibfnamefont {S.}~\bibnamefont
  {Ptasinska}},\ }\href {\doibase 10.1088/1361-6455/ad1745} {\bibfield
  {journal} {\bibinfo  {journal} {J. Phys. B: At. Mol. Opt. Phys.}\ }\textbf
  {\bibinfo {volume} {56}},\ \bibinfo {pages} {245202} (\bibinfo {year}
  {2024})}\BibitemShut {NoStop}%
\end{thebibliography}%

\end{document}